\documentclass[prx,twocolumn,english,superscriptaddress,floatfix,longbibliography]{revtex4-2}

\usepackage[utf8]{inputenc}
\usepackage[T1]{fontenc}
\usepackage{braket,color}
\usepackage{graphicx,enumerate,verbatim,bbold}
\usepackage{amsmath,amssymb,amsthm,float,mathrsfs}
\usepackage{dsfont}
\usepackage[normalem]{ulem}
\usepackage{lmodern}
\usepackage{braket}
\usepackage{multirow}
\usepackage{bbm}
\usepackage{physics}
\usepackage{bm}
\usepackage[normalem]{ulem}
\usepackage{dcolumn}
\newcommand{\bqa}{\begin{eqnarray}}
\newcommand{\eqa}{\end{eqnarray}}

\newcommand{\be}{\begin{equation}}
\newcommand{\ee}{\end{equation}}

\usepackage[caption=false,listofformat=subsimple,labelformat=simple]{subfig}

\begin{document}
\title{Analytically Exact Quantum Simulation of N-Body Interactions via Untunable Decentralized Hamiltonians for Implementing the Toric Code and Its Modifications}

\author{Haochen Zhao}
\email[]{haochen.zhao22@alumni.imperial.ac.uk}
\affiliation{Physics Department, Blackett Laboratory, Imperial College London, Prince Consort Road, SW7 2AZ, United Kingdom}
\author{Florian Mintert}
\affiliation{Physics Department, Blackett Laboratory, Imperial College London, Prince Consort Road, SW7 2AZ, United Kingdom}
\affiliation{Helmholtz-Zentrum Dresden-Rossendorf, Bautzner Landstraße 400, 01328 Dresden, Germany}

\begin{abstract}
We propose a new quantum simulation method for simulating N-body interactions, which are tensor products of N Pauli operators, in an analytically exact manner. This method iteratively attaches many two-body interactions on one two-body interaction to simulate an N-body interaction. Those controlled two-body interactions can be untunable and act only on neighboring spins. The strength difference between controlled and target Hamiltonians is normally not more than one order of magnitude. This simulation is theoretically error-free, and errors due to experimental imperfections are ignorable. A major obstacle to simulating the toric code model and modified toric codes used in topological quantum computation is to simulate N-body interactions. We employ the new quantum simulation method to solve this issue and thus simulate the toric code model and its modifications.
\end{abstract}
\maketitle

\section{Introduction}
\label{sec:introduction}

Topological quantum computation is a method of building a fault-tolerant quantum computer, which uses the braiding of quasi-particles named anyons to perform quantum computation \cite{Pbook}. In particular, the toric code \cite{Ktoric} is one of the commonly used basic models in topological quantum computation, where many modified toric codes \cite{twist,hole} were proposed for implementing a universal quantum computer while minimizing errors. The toric code is a lattice model composed of four-body interactions. The four-body interaction is a Pauli string of length four, which is a tensor product of four Pauli operators. As a side note, we use the terms "Pauli string of length N", "N-body interaction", and "N-body Hamiltonian" to indicate the same object in this paper. However, the experimental realization of topological quantum computation, including the toric code, turns out to be challenging \cite{RMPreviewTQC}. In particular, the complex Hamiltonian of the toric code is a major obstacle to its experimental realization. Although the direct implementation of complex Hamiltonians like the toric code Hamiltonian remains challenging, one can realize those Hamiltonians using quantum simulation. The quantum simulation is to employ experimentally realizable controlled Hamiltonians to simulate a complex target Hamiltonian.

There is an approach in the past paper \cite{hybrid4} for simulating the toric code Hamiltonian by the hybrid quantum simulation \cite{hybrid3}, which is composed of two steps named digital quantum simulation and analog quantum simulation. The whole toric code Hamiltonian can be exactly simulated from four-body interactions using digital quantum simulation. Analog quantum simulation is employed to simulate those four-body interactions from two-body and single-body Hamiltonians. During the analog quantum simulation, sinusoidal controls over two-body interactions are required, with parameters obtained by numerical optimization. In another past paper \cite{nbodyS}, those four-body interactions can be simulated in an analytically exact manner if we have two-body interactions between one reference spin and all other spins under control. Anyons in the toric code can be directly generated and braided using individual Pauli operators, where quantum simulation is not required.

The previous methods \cite{hybrid4,nbodyS} function well when dealing with the toric code as intended. However, if we are interested in simulating not only the original toric code but also the modified toric codes \cite{twist,hole}, the previous methods \cite{hybrid4,nbodyS} will have certain limitations since it is out of the scope of the initial design purpose. The toric code with twists \cite{twist} and the toric code with holes \cite{hole} are two examples of modified toric codes. Five-body interactions are required for simulating the Hamiltonian of the toric code with twists \cite{twist}. Generating anyons in the toric code with holes \cite{hole} needs N-body interactions where N can be larger than ten. Certain applications of the original toric code also require N-body interactions, like the toric code quantum memory \cite{Pbook}. Those Pauli strings of length larger than four cannot be directly simulated by the numerical optimization method \cite{hybrid4}. In an experiment simulating an N-body interaction, one also cannot always ensure that there are two-body interactions between one reference spin and all other spins, so the previous analytical method \cite{nbodyS} is not always available either. A new quantum simulation method which can overcome those limitations is thus required for simulating N-body interactions, which is critical for simulating the toric code and its modifications. Once we can simulate N-body interactions, simulating the toric code and its modifications will be straightforward.

In this paper, we propose a new quantum simulation method named as quantum simulation by attachment. It can simulate N-body interactions in an analytically exact manner without requiring a reference spin. This quantum simulation is composed of a series of iterating steps. In each step, some controlled two-body interactions are attached to one controlled Pauli string to form a longer resulting Pauli string, where the number of bodies in the resulting Pauli string is the sum of the number of bodies in all controlled Pauli strings. The resulting Pauli string in one step is the controlled Pauli string in the next step, and the next step will attach more two-body interactions on this Pauli string to form an even longer Pauli string. This process can be repeated to form an N-body Pauli string from two-body interactions. In this process, a reference spin is not required. Controlled Pauli strings do not need to be tunable, which enables us to implement each step in a short time. In theory, the simulation for a Pauli string with an exponentially increasing number of bodies only requires linearly increasing steps. Under the constraint of experimental setup, in the worst case, the simulation step for simulating a Pauli string with a linearly increasing number of bodies will linearly increase. Even in the worst case, the quantum simulation by attachment is fast enough for most practical purposes. The strength of the resulting N-body interactions is related to the speed of the simulation, so the strength is high enough for most practical purposes since the speed is high. The quantum simulation by attachment is analytically exact, meaning it is error-free in theory. In an experiment, one cannot set parameters to be precisely the same as the theoretical requirement, which will produce certain errors even if there is no error in theory. However, we can show that those experimental errors are small enough to be ignorable.

Quantum simulation by attachment can be directly applied to simulate the toric code and modified toric codes. The digital step of hybrid quantum simulation proposed in the past research \cite{hybrid4,hybrid3} can simplify the question of simulating the whole toric code Hamiltonian into the question of simulating individual four-body interactions, which can then be solved by our method in an analytically exact manner. Anyons of the toric code are generated and braided on the ground state of the toric code Hamiltonian. Generating the ground state can be done using the procedure in the past research \cite{hybrid4} if one can simulate four-body interactions, so we can use quantum simulation by attachment to simulate four-body interactions and thus generate the ground state. Anyons normally can be directly generated and braided by individual Pauli operators. However, in some cases, like the toric code quantum memory \cite{Pbook} and the toric code with holes \cite{hole}, anyons need to be generated and braided by long Pauli strings. Quantum simulation by attachment can be employed to simulate long Pauli strings to generate and braid anyons. Simulating modified toric codes, which are the toric code with twists \cite{twist} and the toric code with holes \cite{hole}, is similar to simulating the original toric code, except Pauli strings longer than four-body interactions are required. The difficulty of simulating long Pauli strings is overcome using quantum simulation by attachment.

This paper is structured as follows. We introduce the quantum simulation by attachment as a new method for simulating three-body and N-body in Sec. \ref{sec:Quantum Simulation by Attachment}. This simulation method is employed to simulate the toric code Hamiltonian in Sec. \ref{sec:Simulating the Toric Code Hamiltonian}. Using this simulation method, the toric code ground state is generated, and anyons are generated and braided on the ground state as demonstrated in Sec. \ref{sec:Ground State and Anyons of the Toric Code}. In Sec. \ref{sec:Simulating Modified Toric Codes}, two modified toric codes, which are the toric code with twists \cite{twist} and the toric code with holes \cite{hole}, are simulated using the quantum simulation by attachment. The strength and experimental errors of the quantum simulation by attachment are discussed in Sec. \ref{sec:Strength and Errors}.

\section{Quantum Simulation by Attachment}
\label{sec:Quantum Simulation by Attachment}

Quantum simulation uses a set of controlled Hamiltonians to generate the same time evolution at certain time points with the target Hamiltonian, where the controlled Hamiltonians are usually single-body or two-body Hamiltonians. In this section, we propose a method, named as quantum simulation by attachment (QSA), for simulating a Pauli string of arbitrary length which is also known as an N-body interaction. The rest of the paper will be based on QSA.

Our approach is based on the relation
\be
\tilde U_1^\dagger \exp(-iH_0T)\ \tilde U_1=\exp(-iH_1T)\ , 
\label{eq:QSA base}
\ee
with
\be
H_1=\tilde U_1^\dagger H_0 \tilde U_1\ ,
\label{eq:QSA base H}
\ee
which is valid for any Hamiltonian $H_0$ and any unitary $\tilde U_1$. This relation implies that the sequence of propagators $\tilde U_1$, $\exp(-i H_0 T)$ and $\tilde U_1^\dagger$ is equivalent to the propagator induced by the transformed Hamiltonian $H_1$. We can treat transformed Hamiltonian $\tilde H$ as the target Hamiltonian and treat Hamiltonians that generate propagators $\tilde U_1$, $\exp(-i H_0 T)$ and $\tilde U_1^\dagger$ as controlled Hamiltonians. Eq. \eqref{eq:QSA base} turns out to precisely satisfy the requirements of being a quantum simulation approach as it generates the same time evolution for controlled and target Hamiltonians. Furthermore, Eq. \eqref{eq:QSA base} can be repeated as
\be
\begin{aligned}
&\tilde U_2^\dagger \tilde U_1^\dagger \exp(-iH_0T) \tilde U_1 \tilde U_2\\
=&\tilde U_2^\dagger \exp(-iH_1T)\ \tilde U_2 \\
=&\exp(-iH_2T)
\end{aligned} 
\label{eq:QSA base further two steps}
\ee
which implies that one can perform a further quantum simulation on the result of a previous quantum simulation to simulate a more complex Hamiltonian iteratively. Obviously, this iteration process can go on and leads to a target Hamiltonian $H_n$ after $n$ steps as
\be
\begin{aligned}
&\tilde U_n^\dagger \ldots \tilde U_2^\dagger \tilde U_1^\dagger \exp(-iH_0T) \tilde U_1 \tilde U_2 \ldots \tilde U_n\\
=&\exp(-iH_nT)
\end{aligned} 
\label{eq:QSA base further}
\ee
It is also worth emphasizing in advance that, in fact, QSA does not need tunable controlled Hamiltonians, whereas single-body or two-body Hamiltonians with fixed strength that can be turned on and off are sufficient. This is a major advantage of QSA since it significantly decreases the experimental difficulty of controlling Hamiltonians to implement QSA. Eqs. \eqref{eq:QSA base} and \eqref{eq:QSA base further} can be rewritten as
\be
e^{i\tilde H_1\tau_1}e^{-iH_0T}e^{-i\tilde H_1\tau_1}=e^{-iH_1T}
\label{eq:QSA simple}
\ee
and
\be
\begin{aligned}
&e^{i\tilde H_n\tau_n} \ldots e^{i\tilde H_2\tau_2}e^{i\tilde H_1\tau_1}e^{-iH_0T}\\
& \times e^{-i\tilde H_1\tau_1}e^{-i\tilde H_2\tau_2} \ldots e^{-i\tilde H_n\tau_n}\\
=&e^{-iH_nT}
\end{aligned}
\label{eq:QSA simple further}
\ee
where all $H_i$ and $\tilde H_i$ are time-independent. Thus, finding controlled Hamiltonians $H_0$ and $\tilde H_i$ in the framework of Eqs. \eqref{eq:QSA simple} and \eqref{eq:QSA simple further}, which can simulate Pauli strings of arbitrary length in an iterative manner, is our central task for building QSA. As a side note, one may notice that in Eq. \eqref{eq:QSA simple} the conjugate unitary $\exp (i\tilde H_1\tau_1)$ requires a Hamiltonian $- \tilde H_1$ which has an opposite sign comparing to $\tilde H_1$ in the original unitary $\exp (i\tilde H_1\tau_1)$, and the same goes in Eq. \eqref{eq:QSA simple further}. This change of sign can be done directly since it is not experimentally difficult because it only requires Hamiltonians with two different fixed strengths during two different time periods instead of a continuous modification of the strength. For our particular chosen controlled Hamiltonians in QSA, there is another approach, where instead of changing the Hamiltonian strength, one can change the time length $\tau_1$ of evolution to generate a conjugate unitary, which will be demonstrated later in this section. In either case, having controlled Hamiltonians with fixed strength that can be turned on and off is sufficient for implementing QSA.

Besides the advantage of not requiring tunable controlled Hamiltonians, another major advantage of QSA is that it is decentralized. It means that QSA does not need two-body interactions between far-apart spins as controlled Hamiltonians. In particular, QSA does not need to have two-body interactions between one particular reference spin and all other spins. We will further explain this decentralizing feature later in the section since it relies on the specific design of the controlled Hamiltonians in QSA. Like the advantage of needing only Hamiltonians with fixed strength, having decentralized controlled Hamiltonians in QSA can also significantly lower the experimental difficulty since connecting all spins to one spin will be more challenging as the number of spins increases.

\subsection{Quantum Simulation by Attachment for Three-body Interactions}

The simplest case of QSA is simulating a three-body Hamiltonian from two-body and single-body Hamiltonians under the framework of Eq. \eqref{eq:QSA simple}. It is also the building block for simulating N-body interactions, which can be done by iteratively repeating three-body QSA as in Eq. \eqref{eq:QSA simple further}. The simulation target, which was labelled by $H_1$ in Eq. \eqref{eq:QSA simple}, of the three-body QSA is given as
\be
H_{t,3}=\sigma_{1}\sigma_{2}\sigma_{3}
\label{eq:QSA three target}
\ee
where $\sigma_{i} \in \{ X_{i}, Y_{i}, Z_{i} \}$ is an arbitrary Pauli operator on spin i. In this subsection, we will first state what controlled Hamiltonians are in QSA and then directly show that those controlled Hamiltonians are indeed capable of simulating a three-body Hamiltonian. After that, we will explain the detailed structure behind those controlled Hamiltonians that support their function. The understanding of this structure provides the ground for generalizing the three-body QSA to N-body QSA, where the N-body QSA will be formally introduced in the next subsection. In later sections, the N-body QSA is a central quantum simulation approach for implementing the toric code and its modifications. In this paper, we use the terms "N-body QSA" and "general QSA" interchangeably.

\begin{figure}[t]
\centering
\includegraphics[width=0.9\columnwidth]{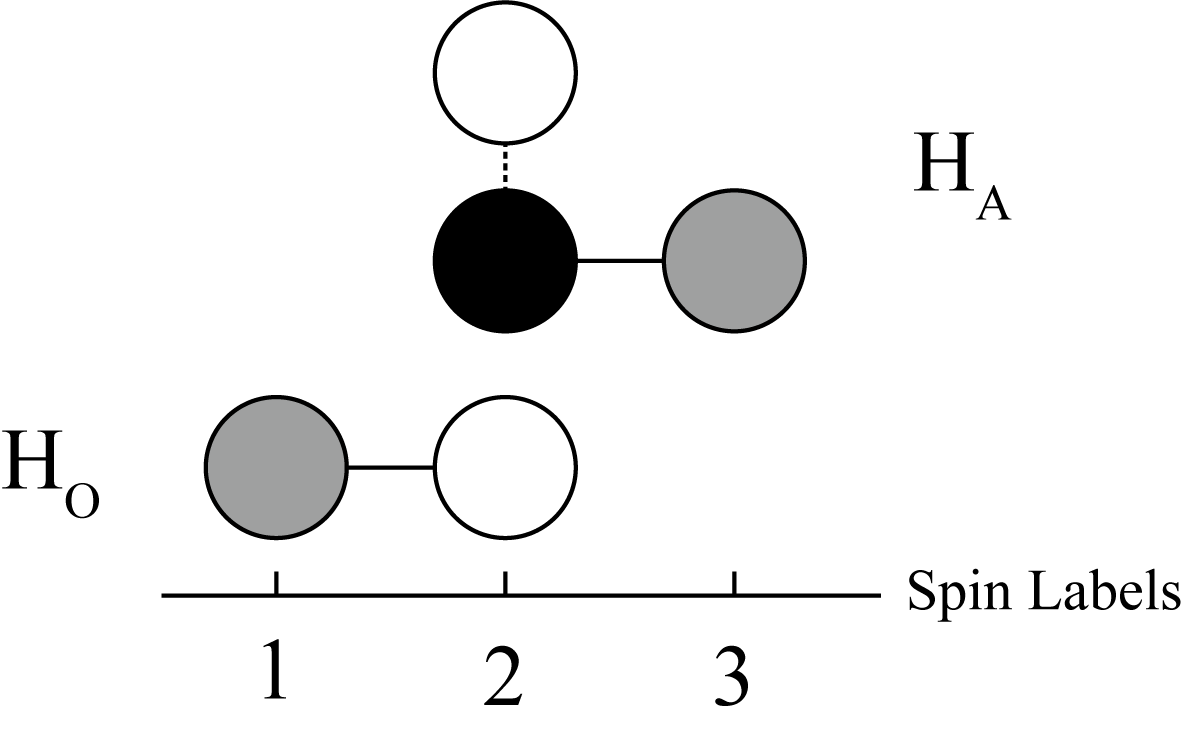}
\caption{Controlled Hamiltonians $H_{A}$ and $H_{O}$ in three-body QSA, where general Pauli operators are represented by grey circles, and connectors are represented by white and black circles. Solid lines between circles indicate tensor products, and the dashed line indicates a summation. $\sigma_{k}$ in $H_{O}$ are artificially chosen to be the same as the white-circle connector in $H_{A}$ for demonstration purposes. The horizontal coordinate labels spin, so circles with the same coordinate are Pauli operators acting on the same spin.}
\label{fig:QSA three control and connectors}
\end{figure}

The controlled Hamiltonian of the three-body QSA is composed of the attachment Hamiltonian $H_{A}$ and the original Hamiltonian $H_{O}$ defined as
\be
\begin{aligned}
& H_{A}=\frac{1}{\sqrt{2}}\left(\sigma_{\alpha} \otimes \mathbb{I}+\sigma_{\beta} \otimes \sigma_3\right) \, , \\
& H_{O}=\sigma_1 \otimes \sigma_{k}
\end{aligned}
\label{eq:QSA three control}
\ee
where $k=\alpha \text{ or } \beta$. $H_{A}$ and $H_{O}$ function as $\tilde H_1$ and $H_0$ in Eq. \eqref{eq:QSA simple} correspondingly, and overall parameters can be put in front of $H_{A}$ and $H_{O}$. $\sigma_{\alpha}$ and $\sigma_{\beta}$ are named as connectors, which are a pair of different single-body Pauli operators acting on the same spin and can be mathematically defined as $\sigma_{\alpha}, \sigma_{\beta} \in \{ X, Y, Z \}, \, \sigma_{\alpha} \neq \sigma_{\beta}$. In fact, any pair of two different Pauli operators on the same spin is a pair of connectors. The concept of connectors is widely used in this paper instead of being limited in Eq. \eqref{eq:QSA three control}. The significance of connectors will be demonstrated when we introduce the structure of three-body QSA later in this subsection. Currently, we are only using this notion to increase the readability of the equation. It is also important to emphasize that $k$ in Eq. \eqref{eq:QSA three control} is indicating one of the connectors instead of indicating the numerical label of the spin, so $\sigma_{k}$ in $H_{O}$ is acting on the same spin with connectors in $H_{A}$. $\sigma_{k}$ can indicate any one connector in the pair $\sigma_{\alpha}$ and $\sigma_{\beta}$ as long as it always indicates the same connector throughout the whole derivation, which is also the point of using a different symbol $k$ instead of $\alpha$ and $\beta$. We can generalize this $k$ symbol to use $k+1$ indicating another connector in the pair, which means $k+1=\alpha$ if $k=\beta$ and $k+1=\beta$ if $k=\alpha$. $k+1$ and $k$ are acting on the same spin, which is the spin acted by $\sigma_{\alpha}$ and $\sigma_{\beta}$, since those $k$ symbols are not indicating the numerical label of the spin as we just mentioned. One can graphically show those controlled Hamiltonians as in Fig. \ref{fig:QSA three control and connectors}, where it is clear that $\sigma_{k}$ in $H_{O}$ is exactly the same with one of the connectors in $H_{A}$, and all connectors in both $H_{A}$ and $H_{O}$ are acting on the same spin.

In three-body QSA, the function of $\tilde U_1$ and $\tilde U_1^\dagger$ in Eq. \eqref{eq:QSA base} is implemented by the attachment propagators $U_{A}$ and $U^{\dagger}_{A}$ correspondingly, which are time evolutions of $H_{A}$ given by
\be
e^{-i \tau \lambda H_A}=\left\{\begin{array}{l}
U_{A}=\frac{i}{\sqrt{2}}\left(\sigma_{\alpha} \otimes \mathbb{I}+\sigma_{\beta} \otimes \sigma_3\right),\\ 
\quad \text {if }  \lambda \tau=\frac{3 \pi}{2}+2 \pi m, \, m \in \mathbb{Z} \\
U^{\dagger}_{A}=\frac{-i}{\sqrt{2}}\left(\sigma_{\alpha} \otimes \mathbb{I}+\sigma_{\beta} \otimes \sigma_3\right),\\ 
\quad \text {if } \lambda \tau=\frac{\pi}{2}+2 \pi m^{\prime},\, m^{\prime} \in \mathbb{Z}
\end{array}\right.
\label{eq:QSA three unitary}
\ee
where $\lambda$ is a controllable real parameter indicating the strength of $H_{A}$, and $m$ and $m^{\prime}$ are arbitrary integers. $m$ and $m^{\prime}$ provides the freedom of obtaining the required attachment propagators from $H_{A}$ by either setting different time length $\tau$ or different strength $\lambda$. If we choose $m=-1$ and $m^{\prime}=0$, we need to have parameters $\lambda \tau=-\pi/2$ for $U_{A}$ and $\lambda \tau=\pi/2$ for $U^{\dagger}_{A}$ in order to obtain required propagators. This condition can be satisfied by using same $\tau$ and opposite $\lambda$ in $U_{A}$ and $U^{\dagger}_{A}$. It physically means that one needs to drive attachment Hamiltonians $H_{A}$ for the same time length but with opposite strength in the experiment. Similarly, if $m=0$ and $m^{\prime}=0$, $\lambda \tau=3\pi/2$ for $U_{A}$ and $\lambda \tau=\pi/2$ for $U^{\dagger}_{A}$ is required, which can be implemented by using the same $\lambda$ and different $\tau$. It means driving $H_{A}$ with the same strength but for different time lengths. Other choices of $m$ and $m^{\prime}$ are also valid, although they are not more efficient than choices just introduced because they need the same or higher strength and time consumption of driving $H_{A}$. In this paper, we will artificially select $m=-1$ and $m^{\prime}=0$ to use as the demonstrative example since they minimize the driving strength and time. Still, all discussions can be applied to all choices of $m$ and $m^{\prime}$ directly or with straightforward modifications.

\begin{figure}[t]
\centering
\includegraphics[width=0.7\columnwidth]{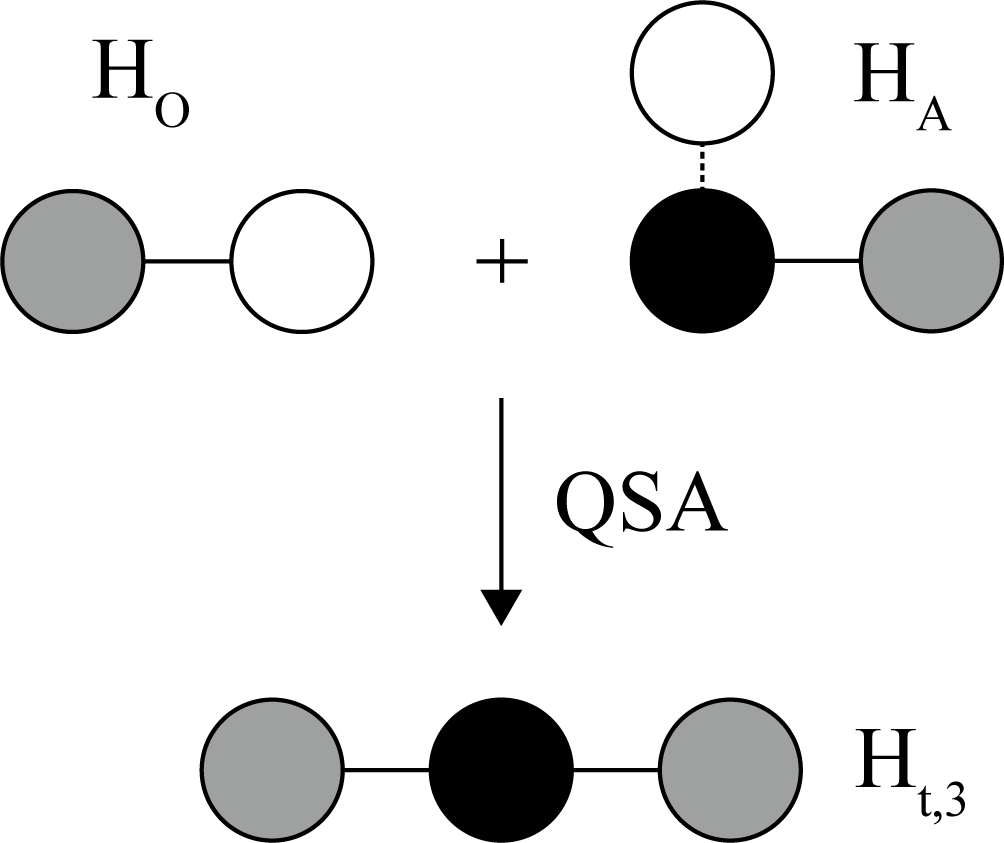}
\caption{QSA for three-body interactions. $H_{t,3}$ is simulated from $H_{O}$ and $H_{A}$.}
\label{fig:QSA three}
\end{figure}

It is mathematically easy to check that
\be
U_{A}H_{O}U^{\dagger}_{A}=\sigma_1 \otimes \sigma_{k+1} \otimes \sigma_3=H_{t,3}.
\label{eq:QSA three H result}
\ee
The last equality is held because connectors are a pair of arbitrary Pauli operators as long as they are different from each other. $\sigma_{k+1}$ is effectively equivalent to $\sigma_{2}$ in $H_{t,3}$, since $\sigma_{k}$ in $H_{O}$ can always be chosen to be not the same with $\sigma_{2}$. Therefore, under the framework of Eq. \eqref{eq:QSA simple}, we have the QSA for three-body interactions being
\be
U_{A} e^{-i t g H_{O}}U^{\dagger}_{A}=e^{-i t g H_{t,3}}
\label{eq:QSA three result}
\ee
where $g$ is a controllable real parameter. It simulates a three-body Hamiltonian from two-body and single-body Hamiltonians and can be graphically represented as Fig. \ref{fig:QSA three}. The QSA for three-body interactions shown in Eq. \eqref{eq:QSA three result} is mathematically straightforward and complete. Now, we will introduce the structure behind Eq. \eqref{eq:QSA three result}. It gives us an intuitive understanding of Eq. \eqref{eq:QSA three result}, which motives and supports the generalization of three-body QSA to N-body QSA.

The whole structure of QSA is built around connectors. In Eqs. \eqref{eq:QSA three H result} and \eqref{eq:QSA three result}, one obvious difference between $H_{O}$ and $H_{t,3}$ in the QSA process is that the connector $\sigma_{k}$ in $H_{O}$ is replaced by the connector $\sigma_{k+1}$ in $H_{t,3}$, which is the key mechanism of QSA. We shall first introduce a reduced version of QSA to illustrate how this connector swapping mechanism works and then show that a simple but non-trivial generalization can give us the whole QSA structure. Another reason for introducing the reduced version of QSA is that this reduced QSA itself is also a component of the N-body QSA, which will be shown in the next subsection.

\begin{figure}[t]
\centering
\includegraphics[width=0.8\columnwidth]{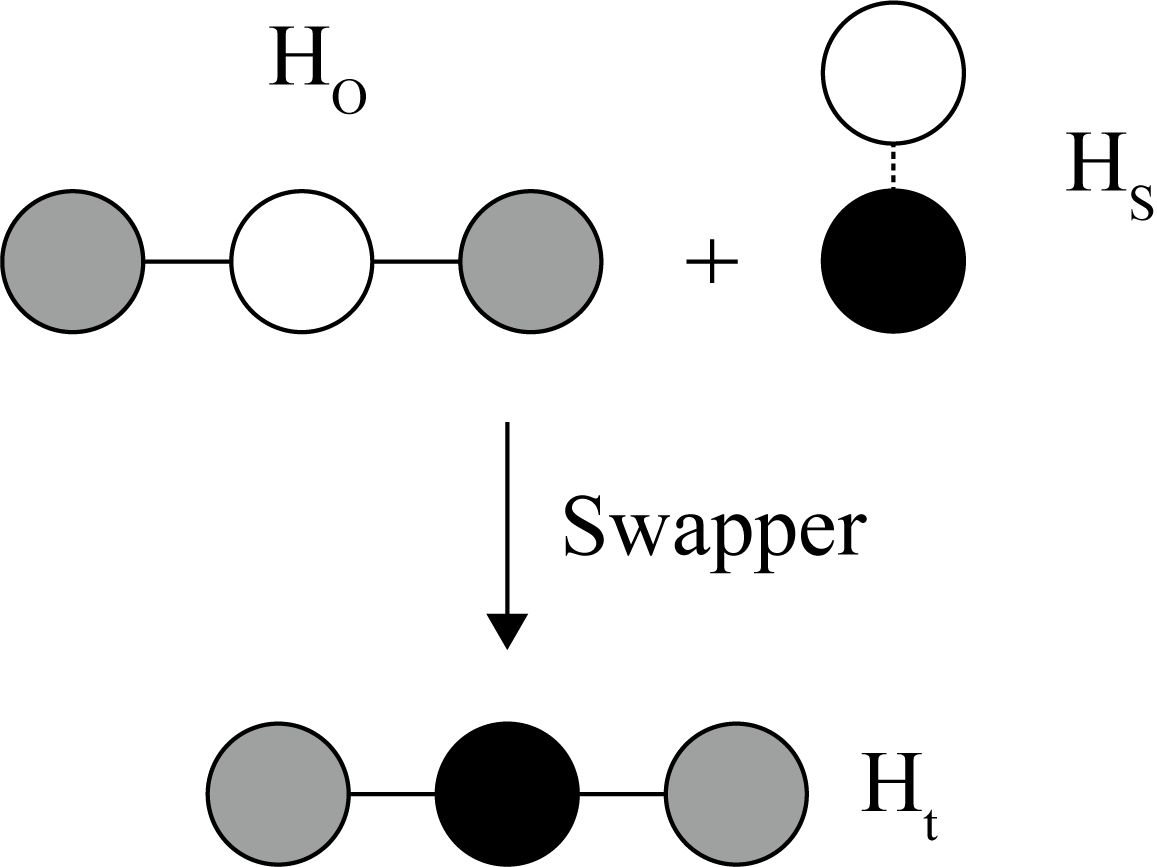}
\caption{Swappers for a three-body interaction. The Pauli operator represented by a white circle in $H_{O}$ is changed to another Pauli operator represented by a black circle.}
\label{fig: QSA three swapper}
\end{figure}

This reduced QSA is named as swappers, where the swapper Hamiltonian $H_{S}$ and the swapper propagators $U_{S}$ and $U^{\dagger}_{S}$ are given as
\be
\begin{aligned}
& H_{S}=\frac{1}{\sqrt{2}}\left(\sigma_{\alpha} +\sigma_{\beta} \right) \, ,\\
& e^{-i \tau \lambda H_S}=\left\{\begin{array}{l}
U_{S}=\frac{i}{\sqrt{2}}\left(\sigma_{\alpha} +\sigma_{\beta}\right),\\
\quad \text {if } \lambda \tau=\frac{3 \pi}{2}+2 \pi m, \, m \in \mathbb{Z}. \\
U^{\dagger}_{S}=\frac{-i}{\sqrt{2}}\left(\sigma_{\alpha} +\sigma_{\beta}\right),\\
\quad \text {if } \lambda \tau=\frac{\pi}{2}+2 \pi m^{\prime},\, m^{\prime} \in \mathbb{Z}.
\end{array}\right.
\end{aligned}
\label{eq:QSA general swapper def}
\ee
Swapper propagators $U_{S}$ and $U^{\dagger}_{S}$ are generated as the time evolution of the swapper Hamiltonian $H_{S}$ in the same way with the attachment propagators $U_{A}$ and $U^{\dagger}_{A}$ being generated from the attachment Hamiltonian $H_{A}$ in Eq. \eqref{eq:QSA three unitary}. We also use $m=-1$ and $m^{\prime}=0$ to minimize the driving strength and driving time. Swappers can change the form of any Pauli operator $\sigma_i$ in a Pauli string Hamiltonian $\sigma_1 \sigma_2 \ldots \sigma_N$ or propagator $\exp (-itg\sigma_1 \sigma_2  \ldots \sigma_N)$ without affecting other Pauli operators as
\be
\begin{aligned}
& U_{S}\sigma_1 \sigma_{k} \sigma_3 \ldots \sigma_N U^{\dagger}_{S}=\sigma_1 \sigma_{k+1}\sigma_3 \ldots \sigma_N \, ,\\
& U_{S} e^{-i t g \sigma_1 \sigma_{k} \sigma_3  \ldots \sigma_N}U^{\dagger}_{S}=e^{-i t g \sigma_1 \sigma_{k+1}\sigma_3 \ldots \sigma_N}.
\end{aligned}
\label{eq:QSA general swapper}
\ee
After applying swappers, connector $\sigma_{k}$ in the original Pauli string Hamiltonian or propagator is replaced by another connector $\sigma_{k+1}$. Since one can choose a pair of arbitrary Pauli operators as connectors, one can always construct the connectors in swapper such that $\sigma_{k}$ in the original Pauli string Hamiltonian is replaced by $\sigma_{k+1}$ which is a Pauli operator with required form. For example, suppose the original Pauli string Hamiltonian is $X_1 Z_2 X_3$, and the target Hamiltonian is $ X_1 Y_2 X_3$. In that case, one can set connects in swappers being $Z_2$ and $Y_2$, which can change the original Pauli string Hamiltonian to the target Hamiltonian by Eq. \eqref{eq:QSA general swapper}. Applying swappers on a three-body Pauli string is shown in Fig. \ref{fig: QSA three swapper}. The idea, or the structure, of this swapper mechanism is firstly to locate one Pauli operator $\sigma_{k}$ in $H_{O}$ and to construct swapper Hamiltonian $H_{S}$ where one of connectors $\sigma_{\alpha}$ and $\sigma_{\beta}$ is the same with $\sigma_{k}$ in $H_{O}$ and another connector is the same with the target Pauli operator, and then to replace this connector $\sigma_{k}$ in $H_{O}$ with anther connector $\sigma_{k+1}$ in the connector pair of $H_{S}$ by Eq. \eqref{eq:QSA general swapper}.

The critical generalization of swappers is in the second step of the structure. Instead of replacing $\sigma_{k}$ with $\sigma_{k+1}$, we can replace $\sigma_{k}$ with $\sigma_{k+1} \otimes \sigma_3$. This generalized swapper is realized by changing the swapper Hamiltonian $H_{S}$ to $\frac{1}{\sqrt{2}}(\sigma_{\alpha} \otimes \mathbb{I}+\sigma_{\beta} \otimes \sigma_3)$, which is identical with changing $H_{S}$ to $H_{A}$. All mathematical procedures of the swapper mechanism are the same with the three-body QSA as shown in Eq. \eqref{eq:QSA three result} and Eq. \eqref{eq:QSA general swapper} excepting the difference between $H_{A}$ and $H_{S}$. Therefore, the generalized swapper is identical to the three-body QSA since $H_{S}$ is changed to $H_{A}$, which eliminates the only difference between swappers and three-body QSA.

One can now make a clear statement about the structure of three-body QSA since it is similar to the swapper structure. Firstly, we locate one Pauli operator $\sigma_{k}$ in $H_{O}$ and construct the attachment Hamiltonian $H_{A}$ based on $\sigma_{k}$. As shown in Eq. \eqref{eq:QSA three control}, $H_{A}$ is composed by a connector $\sigma_{\alpha}$ and a tensor product $\sigma_{\beta} \otimes \sigma_3$, where one of connectors is the same with $\sigma_{k}$ in $H_{O}$ and another connector in $H_{A}$ is the same with $\sigma_{2}$ in $H_{t,3}$. $\sigma_3$ in $H_{A}$ is the same with $\sigma_3$ in $H_{t,3}$. Secondly, we replace the connector $\sigma_{k}$ in $H_{O}$ with $\sigma_{k+1} \otimes \sigma_3$, which is the same with $\sigma_{2} \otimes \sigma_3$, using Eq. \eqref{eq:QSA three result} to produce $H_{t,3}$ as the result. This second step can be thought as swapping connector $\sigma_{k}$ to $\sigma_{k+1}$ where $\sigma_3$ also comes along with $\sigma_{k+1}$. As shown in Fig. \ref{fig:QSA three}, one way to think about the whole three-body QSA mechanism is as follows. We pick a contacting spin in $H_{O}$, and construct a connector pair in $H_{A}$ which acts on this contacting spin. The Pauli operator $\sigma_3$ that exists in $H_{t,3}$ but does not exist in $H_{O}$ is put into $H_{A}$ to form a tensor product $\sigma_{\beta} \otimes \sigma_3$ with one of the connectors. Now we use Eq. \eqref{eq:QSA three result} to swap the connector on the contacting spin in $H_{O}$ from $\sigma_{k}$ to $\sigma_{k+1}$. However, since $\sigma_3$ formed a tensor product with a connector in $H_{A}$, $\sigma_3$ is also swapped into $H_{O}$ by Eq. \eqref{eq:QSA three result}. As spin 3 was not acted by the initial $H_{O}$, the existence of $\sigma_3$ after swapping increases the number of spins of $H_{O}$ by one. It can be described as that three-body QSA attaches one new spin on $H_{O}$. This physical picture will be used throughout this paper. Now, the structure of three-body QSA is sufficiently clear for us, and we are in place to generalize the three-body QSA to N-body QSA, where N-body QSA will be the main result of this paper.

\subsection{General Quantum Simulation by Attachment}
Before formally introducing the general QSA, also known as the N-body QSA, it is worth emphasizing that for the target Pauli string Hamiltonian of N-body QSA, the particular form of those Pauli operators is not important, and the number of bodies is the only relevant factor. For example, target Hamiltonians $XZXZ$, $YYYY$, and $XZZY$ are the mostly equivalent for N-body QSA as their number of bodies is all four. This is done by the application of swappers. As we introduced in Eq. \eqref{eq:QSA general swapper}, the form of any Pauli operator in a Pauli string or a Pauli string propagator can be changed by swappers. It is critical to note that swappers acting on different spins are independent and commutative with each other. One can apply swappers on all spins of an N-body Pauli string or Pauli string propagator simultaneously to change the form of all Pauli operators at once. When the total time consumption of QSA is calculated, the time spending of swappers is unimportant since it always only takes one time step. Given the swapper in the toolbox, the task of simulating a particular N-body Pauli string is simplified to simulating an N-body Pauli string of arbitrary form since the Pauli operators in the string can always be modified to any form in a very short time using swappers.

\begin{figure}[t]
\centering
\includegraphics[width=0.6\columnwidth]{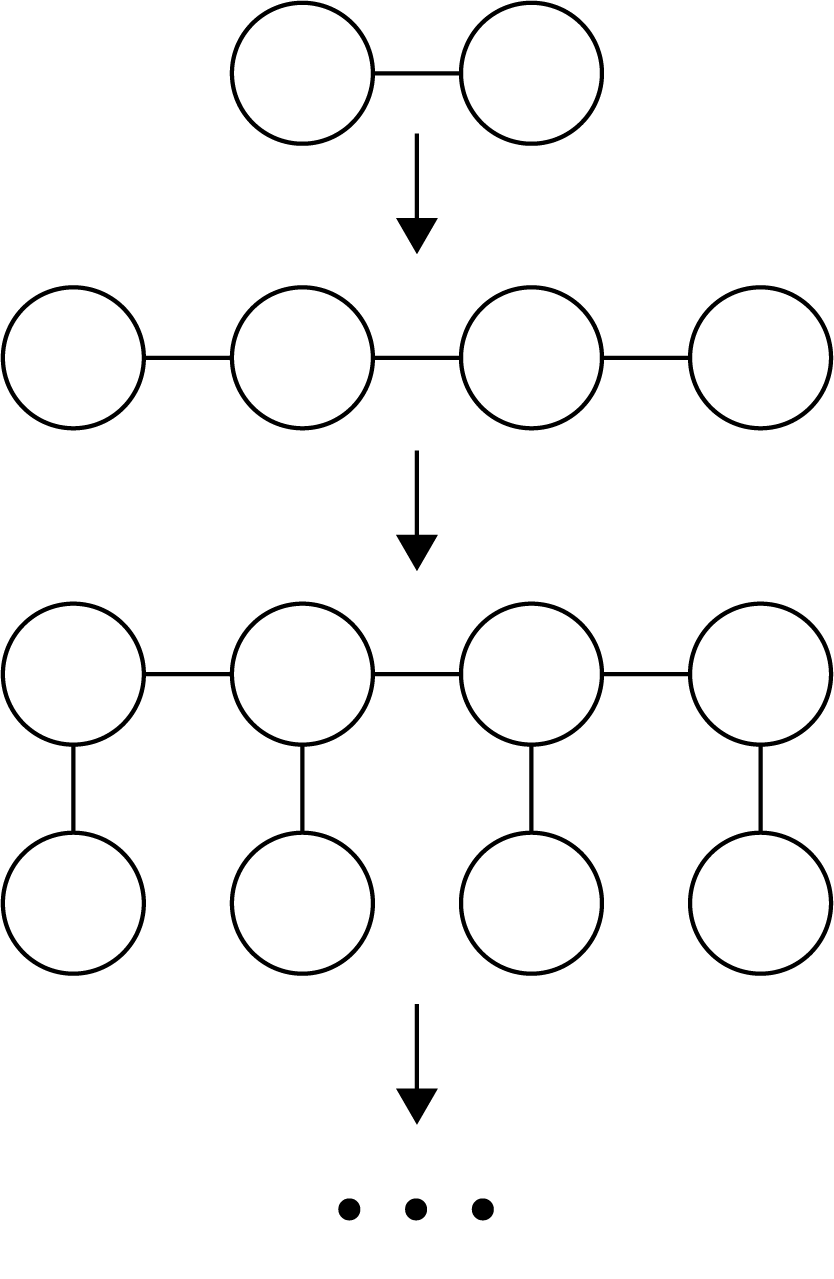}
\caption{General QSA. For each time step, QSA can be applied on all spins of the previous simulation step to double the number of spins in the Pauli string. This process can be repeated to simulate an N-body interaction.}
\label{fig:QSA general max}
\end{figure}

We will now briefly introduce N-body QSA as a generalization of three-body QSA to give an overall picture, and then present the detailed proofs and demonstrations of N-body QSA under the guidance of the overall picture. There are two properties of an individual QSA step, like the three-body QSA, playing a central role when generalizing three-body QSA to N-body QSA. The first property is that any Pauli string $\sigma_1 \sigma_2 \ldots \sigma_N$ can work as $H_O$ in QSA as long as its propagator $\exp (-itg\sigma_1 \sigma_2  \ldots \sigma_N)$ is under control. It means that if we can obtain the propagator of any Pauli string, we can use this Pauli string as $H_O$ in a QSA process. The second property is that all Pauli operators in $H_O$ are able to function as connectors, and multiple three-body QSA can be applied on one $H_O$ simultaneously where each connector in $H_O$ enables one three-body QSA acting on the spin of this connector and each three-body QSA increases the number of body of the target Hamiltonian by one. As shown in Fig. \ref{fig:QSA general max}, those two properties enable us to construct N-body QSA from three-body QSA in an iterative manner. We can start from using a two-body interaction as $H_O$. In the first time step, two three-body QSA can be applied on two spins of the two-body $H_O$, which generates a four-body Pauli string propagator. This four-body Pauli string propagator enables us to use the four-body Pauli string as a new $H_O$. In the second time step, four three-body QSA can be applied on four spins of the four-body $H_O$, which generates an eight-body Pauli string propagator. Those steps can be repeated to simulate a Pauli string of arbitrary length. As a side note, in each time step, applying three-body QSA on all spins of $H_O$ is unnecessary, so a Pauli string of arbitrary length, like a seven-body interaction, can be simulated without issue.

We will demonstrate and prove two properties of QSA along with showing the detailed procedure of N-body QSA, since the N-body QSA provides an excellent example for introducing properties of QSA and those properties support the validity of N-body QSA. The first time step of N-body QSA is a direct generalization of three-body QSA, so we first look into three-body QSA. In Eq. (\ref{eq:QSA three control}) of three-body QSA, $H_A$ is independent and thus commute with $\sigma_1$ in $H_O$, so Eq. (\ref{eq:QSA three H result}) also can be written as
\be
U_{A}\sigma_1 \sigma_{k} U^{\dagger}_{A}=\sigma_1 (U_{A} \sigma_{k} U^{\dagger}_{A})=\sigma_1 \sigma_{k+1} \sigma_3 \, .
\label{eq:QSA general independence}
\ee
This observation implies that the $\sigma_1$ in $H_O$ can be replaced by any Hamiltonian, including an N-body Pauli string, as long as it commutes with $H_A$. To avoid confusion, we name any Hamiltonian that replaces $\sigma_1$ in $H_O$, including $\sigma_1$ itself, as the stable Hamiltonian in $H_O$. This replacement property enables a Pauli string of arbitrary length to be a valid $H_O$, since in order to construct an N-body $H_O$ one can use an (N-1)-body Pauli string as the stable Hamiltonian. Besides that, in Eq. \eqref{eq:QSA three result}, we can find that only the propagator $\exp (-i t g H_{O})$ of $H_O$, instead of $H_O$ itself, is required in the three-body QSA process, which means we can perform three-body QSA without using $H_O$ itself if we have the propagator $\exp (-i t g H_{O})$ under control. It is still true if we use an arbitrary N-body Pauli string as the $H_O$, since only the stable Hamiltonian of $H_O$ is modified but the stable Hamiltonian does not affect the mechanism of Eq. \eqref{eq:QSA three result} as shown in Eq. \eqref{eq:QSA general independence}. The first property of QSA is thus proofed, which is that $H_O$ can be an arbitrary N-body Pauli string $\sigma_1 \sigma_2 \ldots \sigma_N$, and the propagator $\exp (-itg\sigma_1 \sigma_2  \ldots \sigma_N)$ instead of $H_O$ itself is required in a QSA step.

In the case of using a Pauli string as the stable Hamiltonian, the distinction between the stable Hamiltonian and the connector in $H_O$ is highly artificial. The connector as an arbitrary Pauli operator is identical to any individual Pauli operator in the stable Hamiltonian. One is free to choose any Pauli operator in $H_O$ as the connector and the rest as the stable Hamiltonian, which is the first half of the second property. The second half of the second property states that multiple QSA can be simultaneously applied on the same $H_O$ using different connectors, which is more tricky. It is clearer to introduce the proof of the second half after showing the N-body QSA, so we are now back to the first time step of N-body QSA. 

The first time step of N-body QSA is a four-body QSA composed of two three-body QSA, but it also can be considered as applying one extra three-body QSA on the result of another three-body QSA. We will initially follow the second perspective of the first time step and then show that those two perspectives are equivalent. $H_{t,3}=\sigma_1\sigma_{2}\sigma_{3}$ as the simulation result of three-body QSA can be treated as a new original $H_O$ due to the first property of QSA. We choose $\sigma_1$ as the connecter and $\sigma_{2}\sigma_{3}$ as the stable Hamiltonian, so the new attachment Hamiltonian $H_{A,1}$ for this $H_O$ is
\be
H_{A,1}=\frac{1}{\sqrt{2}}\left(\sigma_{\alpha,1} \otimes \mathbb{I}+\sigma_{\beta,1} \otimes \sigma_0\right)
\label{eq:QSA general HA}
\ee
which is independent with $\sigma_{2}\sigma_{3}$ as required. $\sigma_{\alpha,1}$ and $\sigma_{\beta,1}$ are connecters on spin 1, where we use an integer 1 besides $\alpha$ and $\beta$ to indicate the spin acted by the connector. We can thus simulate a four-body interaction from a three-body interaction following the same QSA procedure with Eq. \eqref{eq:QSA three result}. Since $H_O$ here is the simulation result of a previous QSA, this four-body interaction is simulated from two-body interactions by two QSA processes, which can be expressed as
\be
U_{A,1}U_{A,2} e^{-i t g \sigma_{1} \sigma_{2}}U^{\dagger}_{A,2}U^{\dagger}_{A,1}=e^{-i t g \sigma_{0} \sigma_{1}^\prime \sigma_{2}^\prime \sigma_{3}} \, .
\label{eq:QSA general four}
\ee
We relabeled the $H_A$ of three-body QSA in Eq. (\ref{eq:QSA three control}) as $H_{A,2}$. Propagators $U_{A,1}$ and $U_{A,2}$ are the time evolutions of $H_{A,1}$ and $H_{A,2}$ correspondingly given by Eq. \eqref{eq:QSA three unitary}. We also relabeled the $\sigma_{k}$ in the $H_O$ of Eq. \eqref{eq:QSA three result} as $\sigma_{2}$, and thus label swapped connectors, which were $\sigma_{1}$ and $\sigma_{2}$ before QSA, as $\sigma_{1}^\prime$ and $\sigma_{2}^\prime$ correspondingly. However, since all $\sigma_{i}$ and $\sigma_{i}^\prime$ are arbitrary Pauli operators excepting that any specific pair of $\sigma_{i}$ and $\sigma_{i}^\prime$ must be composed by different Pauli operators, there is no fundamental difference between $\sigma_{i}$ and $\sigma_{i}^\prime$ as long as they are not being compared with each other. From now, we will ignore the difference between $\sigma_{i}$ and $\sigma_{i}^\prime$ if it is not relevant in the context, and use $\sigma_{i}$ to indicate both $\sigma_{i}$ and $\sigma_{i}^\prime$.

Since $H_{A,1}$ is independent with $H_{A,2}$, the pair of propagators $U_{A,1}$ and $U_{A,2}$, as well as the $U^{\dagger}_{A,1}$ and $U^{\dagger}_{A,2}$ pair, are independent and commutative. It means QSA processes acting on $\sigma_{1}$ and $\sigma_{2}$ can be done simultaneously. To simplify the demonstration, we count the time consumption of one three-body QSA process described in Eq. (\ref{eq:QSA three result}) as taking one time step. Since two QSA processes corresponding to $U_{A,1}$ and $U_{A,2}$ can be done simultaneously, simulating a four-body interaction from two-body interactions by QSA, as shown in Eq. \eqref{eq:QSA general four}, takes only one time step. This is the first time step of N-body QSA and an example of the second property of QSA, where two three-body QSA are acting on two different connectors simultaneously. In order to show that this property is generally valid, we now move to the next time step of N-body QSA to have a clearer context about how to state a general attachment Hamiltonian and its corresponding three-body QSA.

The four-body Pauli string $\sigma_{0} \sigma_{1} \sigma_{2} \sigma_{3}$ as the simulation result of the first time step of N-body QSA is also a valid $H_O$ for further QSA steps. It means QSA can also be applied on $\sigma_{0} \sigma_{1} \sigma_{2} \sigma_{3}$ to simulate a longer Pauli string, where any $\sigma_{i}$ in $\sigma_{0} \sigma_{1} \sigma_{2} \sigma_{3}$ can be treated as a connector and others can be treated as the stable Hamiltonian. The fact that QSA need to be applied on many different spins motivates us to construct a more general attachment Hamiltonian $H_{A,i}$, which is the attachment Hamiltonian of QSA acting on spin i, as
\be
H_{A,i}=\frac{1}{\sqrt{2}}(\sigma_{\alpha,i} \otimes \mathbb{I}+\sigma_{\beta,i} \otimes \sigma_{m})
\label{eq:QSA general general attachment}
\ee
where $\sigma_{m}$ is a Pauli operator acting on spin m. There is a restriction that the three-body QSA employing $H_{A,i}$ must not act on an $H_O$ containing spin m. It means $\sigma_{m}$ always provides an extra spin for $H_O$. We construct one $H_{A,i}$ for each connector of an $H_O$ in one time step of N-body QSA, where $\sigma_{m}$ in different $H_{A,i}$ is artificially chosen to be acting on different spins with each other. All $H_{A,i}$ for the same $H_O$ are thus commutative with each other because connectors and $\sigma_{m}$ in different $H_{A,i}$ are acting on different spins. It means that all three-body QSA acting the same $H_O$ can be done simultaneously, which is the second property of QSA. As a particular case, all four three-body QSA acting on $\sigma_{0} \sigma_{1} \sigma_{2} \sigma_{3}$ can be done simultaneously. We can simulate an eight-body interaction using four three-body QSA in the second time step. This means that one can simulate an eight-body interaction from two-body interactions using N-body QSA in two time steps.

However, as we will prove soon, attachment Hamiltonians of the second step QSA acting on $\sigma_{0} \sigma_{1} \sigma_{2} \sigma_{3}$ do not commute with all attachment Hamiltonians of the first step QSA acting on $\sigma_{1} \sigma_{2}$ shown in Eq. \eqref{eq:QSA general four}. For example, if we pick $\sigma_{0}$ as the connecter, the attachment Hamiltonian $H_{A,0}^\prime=\frac{1}{\sqrt{2}}(\sigma_{\alpha,0} \otimes \mathbb{I}+\sigma_{\beta,0} \otimes \sigma_{-1})$ does not commute with $H_{A,1}$ in Eq. \eqref{eq:QSA general four}. The noncommutativity between some attachment Hamiltonians of the second step QSA acting on $\sigma_{0} \sigma_{1} \sigma_{2} \sigma_{3}$ and some attachment Hamiltonians of the first step QSA acting on $\sigma_{1} \sigma_{2}$ means that those second step QSA cannot be done simultaneously with the first step QSA. That is why those QSA must require two time steps instead of one. 

We can now provide clear proof of the noncommutativity between the second step and the first step attachment Hamiltonians. Since this noncommutativity is valid for any pair of two succeeding time steps in N-body QSA, we name those time steps as the further and the previous time steps. In the N-body QSA process, the further step is done after the previous step. For the pair of the second and the first steps, the second step is the further step, and the first step is the previous time step. Noncommutativity is the major source of time consumption in N-body QSA as it requires N-body QSA to take multiple time steps, which is why it deserves a more detailed discussion. It is important to emphasize that the noncommutativity only holds between certain but not all attachment Hamiltonians of further and previous QSA. The statement that we really want to prove is that the further QSA always needs one more time step.

To illustrate the proof, we shall firstly consider the case of having further QSA acting on $\sigma_{0} \sigma_{1} \sigma_{2} \sigma_{3}$ where $\sigma_{0} \sigma_{1} \sigma_{2} \sigma_{3}$ is the result of a previous QSA acting on $\sigma_{1} \sigma_{2}$. This means that the second time step of N-body QSA is the further QSA, and the first time step is the previous QSA. Attachment Hamiltonians in the further QSA are labelled as $H_{A,i}^\prime$ and attachment Hamiltonians in the previous QSA are labelled as $H_{A,i}$. $H_{A,1}$ acts on spin 0 and spin 1, and $H_{A,2}$ acts on spin 2 and spin 3. $H_{A,0}^\prime$ and $H_{A,1}^\prime$ only have common spin with $H_{A,1}$, and $H_{A,2}^\prime$ and $H_{A,3}^\prime$ only have common spin with $H_{A,2}$. The set of $H_{A,0}^\prime$, $H_{A,1}^\prime$ and $H_{A,1}$ is independent and commutative with the set of $H_{A,2}^\prime$, $H_{A,3}^\prime$ and $H_{A,2}$, but Hamiltonians do not commute within a set. The first set can be taken as an example for now. It can be easily checked that both $H_{A,0}^\prime$ and $H_{A,1}^\prime$ do not commute $H_{A,1}$ via direct calculation of commutators $[H_{A,0}^\prime,H_{A,1}]$ and $[H_{A,1}^\prime,H_{A,1}]$, so the further QSA using $H_{A,0}^\prime$ or $H_{A,1}^\prime$ cannot be done simultaneously with the previous QSA using $H_{A,1}$. The same discussion is also valid for the set of $H_{A,2}^\prime$, $H_{A,3}^\prime$ and $H_{A,2}$.

However, $H_{A,0}^\prime$ and $H_{A,1}^\prime$ are naturally commutative with any $H_{A,i}$ other than $H_{A,1}$ since they acting on different spins, which is another way of expressing different sets are independent. It will not damage the statement that the further QSA always needs one more time step because the existence of $H_{A,1}$ in the previous QSA will force the further QSA to use one more time step due to the noncommutativity related to the $H_{A,1}$. If we do not employ $H_{A,1}$ in the previous QSA, the noncommutativity related to the $H_{A,2}$ can still ensure that the further QSA needs one more step. If both $H_{A,1}$ and $H_{A,2}$ are not employed in the previous QSA, it is the same as doing nothing in the previous QSA, which violates the background assumption of having two QSA steps.

Furthermore, $H_O$ of the further QSA can be an N-body Pauli string due to the first property of QSA, where this N-body Pauli string can be assumed to be the simulation result of a previous QSA applied on an $N/2$-body Pauli string. In this more general circumstance, the proof used in the four-body case is still valid. Each attachment Hamiltonian in the previous step can form a set with two attachment Hamiltonians in the further step, which have common spins with this attachment Hamiltonian in the previous step, in the same manner as constructing sets in the four-body case. Just like in the four-body case where the noncommutativity within each set demands one extra time step, in the general case the noncommutativity within each set also requires one extra time step. The fact that there are multiple sets will not make the further QSA take more than one extra time step because different sets are independent and commutative. Therefore, in the general N-body Pauli string case, the further QSA indeed only requires one extra time step and no more, which is the end of the proof.

As shown in Fig. \ref{fig:QSA general max}, the procedure of applying further QSA on the simulation result of the previous QSA can be iterated since we have proven that two properties of QSA are generally valid. Starting from a two-body Hamiltonian, we can apply QSA on both spins of this two-body Hamiltonian to simulate a four-body Hamiltonian in the first time step. In the second time step, we can apply QSA on all four spins of the four-body Hamiltonian, which is the simulation result of the first time step, to simulate an eight-body Hamiltonian. In the third time step, QSA can be applied on all eight spins of the eight-body Hamiltonian to simulate a sixteen-body Hamiltonian, and so on. This iterated QSA process is the general QSA, also known as the N-body QSA, which can simulate an arbitrary N-body interaction from two-body interactions. In this paper, we also use the term QSA to refer to general QSA since three-body and four-body QSA can be treated as particular examples of general QSA with one time step. With linearly increasing time steps $t$, the number of spins $N$ in the resulting Pauli string will increase exponentially in general QSA where $N=2^{t+1}$, since $N$ will be doubled in each time step. It is not essential to apply three-body QSA to all spins in a time step. This is how we could simulate Pauli strings with the number of bodies $N$ being not exactly integers $2^{t+1}$. A Pauli string with $N$ being between $2^{t}$ and $2^{t+1}$ can be simulated by the general QSA with $t$ time steps, where in the last step we only apply three-body QSA to $N - 2^{t}$ spins instead of all spins. As a side note, if one wants to have a Pauli string containing identity operators in some specific spins, like $\sigma_{0} \sigma_{1} \mathbb{I}_{2} \sigma_{3}$, it can be done by not driving those spins during the whole N-body QSA process. 

The specific form of the resulting Pauli string can be modified to any required form by simultaneously applying swappers on spins that need to be edited. This modification always takes only one time step as we already knew, but now, with the knowledge of the general QSA, we can further state that the time step used by swappers normally can be much shorter in an experiment than other time steps in the general QSA. The reason is that swappers comprise only single-body Hamiltonians instead of two-body Hamiltonians used in attachment Hamiltonians of other time steps. Single-body Hamiltonians are easier to control and thus can be stronger than two-body Hamiltonians. Since the products $\lambda \tau$ of strength and time are set to fixed values when generating propagators from Hamiltonians in Eq. \eqref{eq:QSA three unitary} and Eq. \eqref{eq:QSA general swapper def}, swapper propagators can take shorter time than attachment propagators. Therefore, we can usually ignore the time consumption of swappers when counting the total time consumption of the general QSA. It is also important to emphasize that other individual steps in the general QSA are also very fast, even though they are usually slower than swappers. The reason is that we only use untunable two-body interactions in the attachment Hamiltonians, which are normally easy to construct in an experiment. The rest is the same argument as the swapper case. Simplicity in constructing Hamiltonians likely enables the Hamiltonian to have a high strength. Due to Eq. \eqref{eq:QSA three unitary}, attachment propagators can be obtained in a short time using attachment Hamiltonians with high strength. It means each time step of the general QSA is expected to be very short.

\begin{figure}[t]
\centering
\includegraphics[width=0.9\columnwidth]{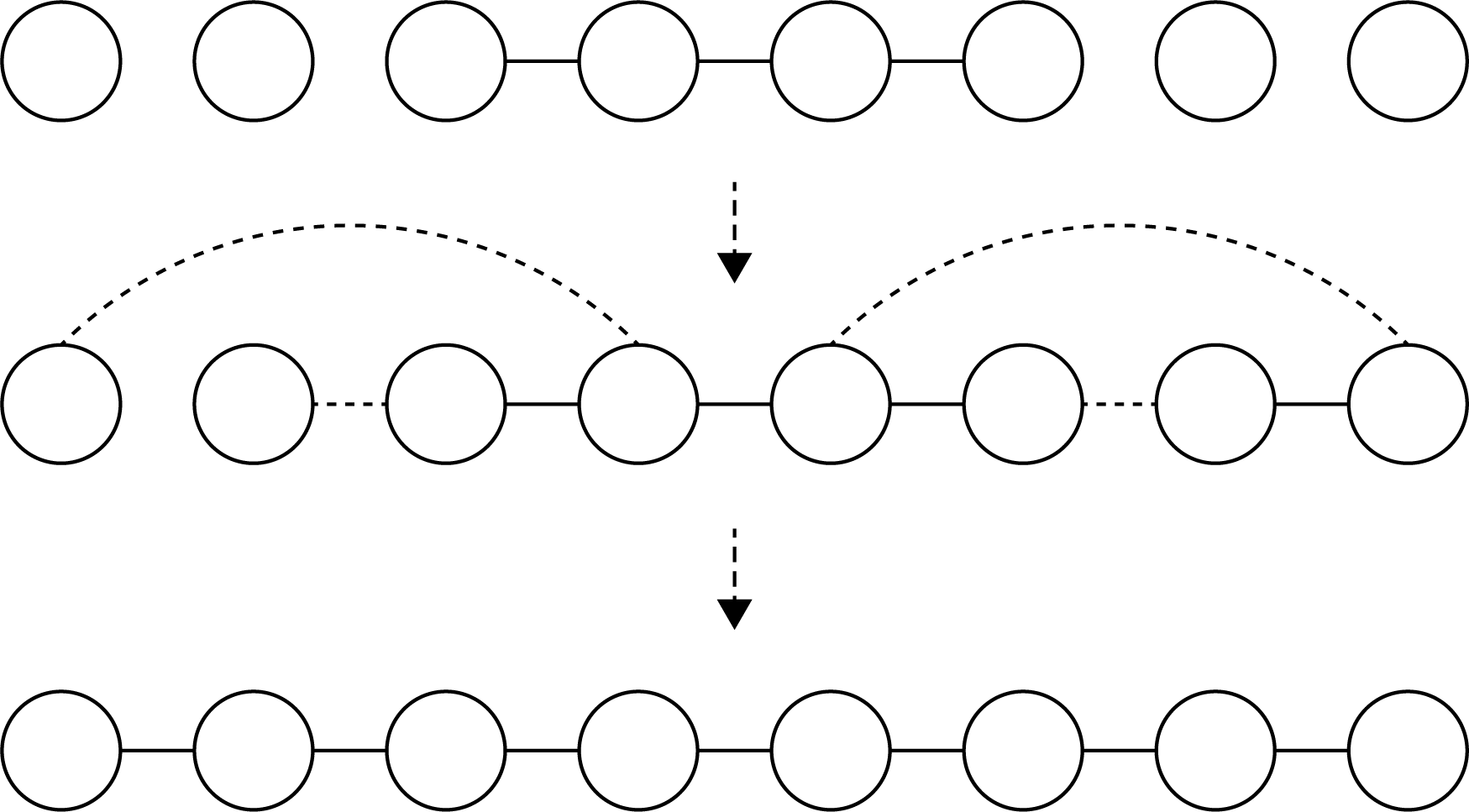}
\caption{General QSA with exponentially increasing spin number for simulating a Pauli string line. A step of the general QSA, which simulates eight-body interaction from four-body interaction, is split into two small steps in this figure, where the dashed arrow is used to emphasize that one arrow is not a full QSA step. The first small step builds two-body interactions indicated by dashed lines, which are employed in attachment Hamiltonians. The second small step uses those attachment Hamiltonians to simulate the eight-body interaction.}
\label{fig:QSA general line unable}
\end{figure}

\begin{figure}[t]
\centering
\includegraphics[width=\columnwidth]{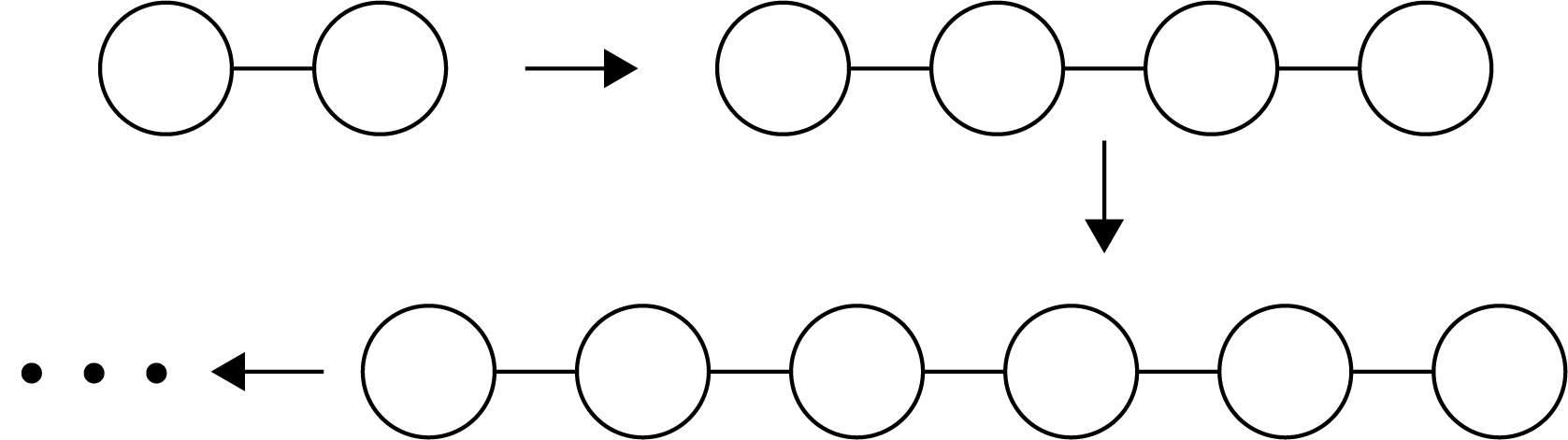}
\caption{General QSA for a Pauli string line, where one can only perform QSA on the endpoints of the Pauli string line.}
\label{fig:QSA general line}
\end{figure}

The general QSA with an exponentially increased number of spins is the fastest general QSA, but one may not always be able to do it in experiments. The key issue is that one may not always be able to construct two-body interactions between any spins in an experiment. However, one major advantage of the general QSA is that we can modify the general QSA to overcome this difficulty. In each step of the general QSA, one can apply three-body QSA to certain but not all spins. Pairs of spins that cannot be connected by two-body interaction in a particular experiment can be avoided in all individual steps of the general QSA, which means attachment Hamiltonians employing those two-body interactions are not used in this version of the general QSA. In particular, one of the worst but common cases in an experiment is that one wants to simulate a Pauli string where spins are spatially placed in a line, but one only has two-body interactions between nearest neighbor spins. It is a common situation in the toric code \cite{Ktoric} and its modifications \cite{twist,hole} as we will see in the following sections. As shown in Fig. \ref{fig:QSA general line unable}, the general QSA with exponentially increasing spin number cannot handle this case. To double the spin number, we must have two-body interactions between the spin in the middle of the line and the spin at the endpoints, which are not nearest neighbor spins if the line is longer than four spins. We can solve this problem by only applying QSA to two spatial endpoints of the Pauli string in each time step of the general QSA as in Fig. \ref{fig:QSA general line}. We can still simulate an arbitrary N-body interaction using this version of the general QSA, although at a relatively slower speed. With linearly increasing time steps $t$, the number of spins $N$ here will increase linearly as $N=2(t+1)$. It is normally fast enough for practical usage since each time step is usually very short. The theoretical worst case is that we can apply three-body QSA to only one spin in each time step. In this case, $N$ also increases linearly as $N=t+2$, so general QSA is fairly fast even in the worst case possible.

One way to express this major advantage of the general QSA we just mentioned is that controlled Hamiltonians in the general QSA are decentralized. In each iterating step of the general QSA, three-body QSA can be applied on any spin of the Pauli string since all spins are capable of being connectors, so the whole general QSA does not rely on any specific spin. Different experimental setups allow two-body interactions to be constructed on different pairs of spins. Decentralized controlled Hamiltonians enable us to perform general QSA under those variant experimental restrictions. As a particular example, Fig. \ref{fig:QSA general line unable} and Fig. \ref{fig:QSA general line} demonstrated the value of having decentralized controlled Hamiltonians, where simulating a spatially linear Pauli string using only nearest neighbor is impossible without the decentralization feature of the general QSA. This advantage is especially beneficial when simulating the toric code and its modifications, which will be introduced in later sections.

Before moving to the application of QSA on the toric code and its modifications, we shall briefly introduce the strength and errors of QSA, although leaving further discussions in Sec. \ref{sec:Strength and Errors}. As shown in Eq. (\ref{eq:QSA three result}) and Eq. (\ref{eq:QSA general four}), $H_O$ and $H_t$ have the same combined coefficient $t g$ coefficient in QSA. This means that $t g$ is preserved during the whole QSA process regardless of how many layers of $U_A$ are applied to $H_O$. Although the product of coefficients $t g$ is preserved, individual coefficients $t$ and $g$ are not necessarily the same for $H_O$ and $H_t$. In Eq. (\ref{eq:QSA three result}) and Eq. (\ref{eq:QSA general four}), if we assume the total time evolution of controlled Hamiltonians on the left-hand side of equations have the same time spending with the time evolution of the target Hamiltonian on the right-hand side of equations, $t$ and $g$ of $H_t$ actually did not represent the time and the strength of $H_t$ corresponding. The time consumption of $H_t$ must be longer than $t$ of $H_O$ since the time consumption of $H_t$ is assumed to be the same as the total time consumption of attachment propagators and $H_O$. In order to solve this issue, we relabel $t g$ in $H_t$ as $t^{\prime} g^{\prime}$ where $t^{\prime}$ is the time consumption of $H_t$ and $g^{\prime}$ is the strength of $H_t$, while keeping $t g$ in $H_O$ unchanged. The preservation of $t g$ in QSA can be ensured by setting the relation
\be
t g= t^{\prime} g^{\prime}.
\label{eq:QSA strength time relation}
\ee
As the $t^{\prime}$ is larger than $t$, Eq. \eqref{eq:QSA strength time relation} claim that $g^{\prime}$ is smaller than $g$. If we consider QSA as a process of changing $H_O$ to $H_t$ using many layers of attachment propagators, the difference between $t^{\prime}$ and $t$ is effectively a measure of the speed of QSA since this difference is the time consumption of all attachment propagators. As QSA is expected to be fast, the difference between $t^{\prime}$ and $t$ should be small, so the difference between $g^{\prime}$ and $g$ is also small due to Eq. \eqref{eq:QSA strength time relation}. As shown in Sec. \ref{sec:Strength and Errors}, the difference between $g^{\prime}$ and $g$ is normally not more than one order of magnitude in the QSA application on the toric code and its modifications, which is usually acceptable in an experiment.

The general QSA does not produce any error in theory since the general QSA is analytically exact, as we have seen in this section. However, in an experiment, one is not able to set parameters exactly the same as the theoretical requirement. In Sec. \ref{sec:Strength and Errors}, we will check the error from experiment imperfection in the general QSA itself as well as in the QSA applications in the toric code and its modifications to show that this error is sufficiently small and can be ignored.

\section{Simulating the Toric Code Hamiltonian}
\label{sec:Simulating the Toric Code Hamiltonian}

\begin{figure}[t]
\centering
\includegraphics[width=0.5\columnwidth]{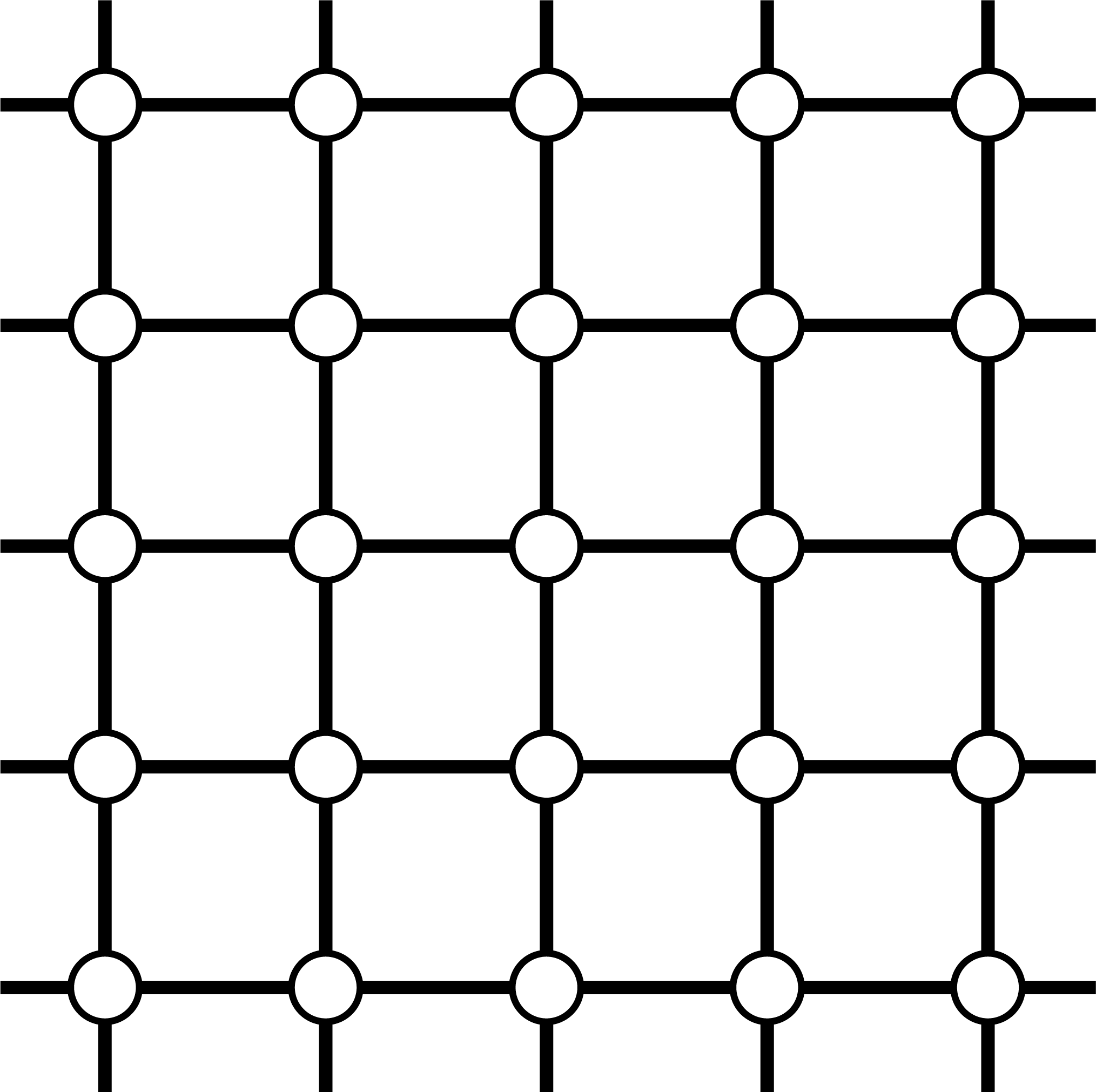}
\caption{A square lattice where the dots are spin 1/2 particles. The toric code Hamiltonian is defined on this lattice.}
\label{fig:toric wen}
\end{figure}

We can now apply QSA to simulate the toric code \cite{Ktoric} and its modifications \cite{twist,hole}. We will start by simulating the Hamiltonian of the toric code in this section. The toric code Hamiltonian has two equivalent models \cite{Ktoric,Wtoric}, where, inspired by the past research on simulating the toric code \cite{hybrid4}, Wen’s model \cite{Wtoric} will be used now, given as
\be
\begin{aligned}
& H_w=-J \sum_{i, j} P_{i, j} \, , \\
& P_{i, j}=X_{i, j} Z_{i, j+1} Z_{i+1, j} X_{i+1, j+1} \, .
\end{aligned}
\label{eq:toric wen}
\ee
It is defined on a square lattice composed of spin 1/2 particles as shown in Fig. \ref{fig:toric wen}. Each $P_{i, j}=X_{i, j} Z_{i, j+1} Z_{i+1, j} X_{i+1, j+1}$ is defined as four Pauli operators acting on four spins of one plaquette of the lattice, where i and j labelling the spins of action. The plaquette operator $P_{i, j}$ is summed over the whole lattice with a coefficient $J$ to generate the toric code Hamiltonian $H_w=-J \sum_{i, j} P_{i, j}$. This Hamiltonian is our simulation target in this section.

From now on, we will assume that we have two-body interactions between nearest neighbor and next-nearest neighbor spins in the lattice as well as single-body Hamiltonians on all spins under control since it is normally possible in an experiment. For example, we can have nearest neighbors $\sigma_{i, j}\sigma_{i, j+1}$ and next-nearest neighbors $\sigma_{i, j}\sigma_{i+1, j+1}$ as our controlled Hamiltonians.

To simulate the toric code Hamiltonian, we will follow mostly the same procedure with a past paper \cite{hybrid4}, but one central step will be replaced by the QSA. In the past paper \cite{hybrid4}, as well as in our research, the hybrid quantum simulation \cite{hybrid3,hybrid4} serves as the general framework for simulating the toric code. The hybrid quantum simulation is composed of an analog quantum simulation and a digital quantum simulation. The digital quantum simulation can simulate the whole toric code Hamiltonian from individual plaquette operators $P_{i, j}$ \cite{hybrid4}, which will be introduced later in this section. The analog quantum simulation is to simulate those individual plaquette operators $P_{i, j}$ from single-body and two-body Hamiltonians. In the past research \cite{hybrid4}, individual plaquette operators $P_{i, j}$ are firstly simulated from tunable single-body and two-body controlled Hamiltonians using the analog quantum simulation, then the whole toric code Hamiltonian is simulated from those $P_{i, j}$ using the digital quantum simulation. In our research, we follow the same procedure of firstly simulating $P_{i, j}$ and then simulating the whole toric code Hamiltonian. However, in the first step, we replace the analog quantum simulation with four-body QSA so that we could employ only untunable controlled Hamiltonians. It significantly decreases the experimental difficulty and thus increases the speed of simulation and the strength of the target Hamiltonian. Now we will firstly illustrate how to simulate $P_{i, j}$ from two-body interactions using QSA, then demonstrate that the whole toric code Hamiltonian can indeed be simulated from those  $P_{i, j}$, which are the simulation results of QSA, using the digital quantum simulation.

\begin{figure}[t]
\centering
\includegraphics[width=\columnwidth]{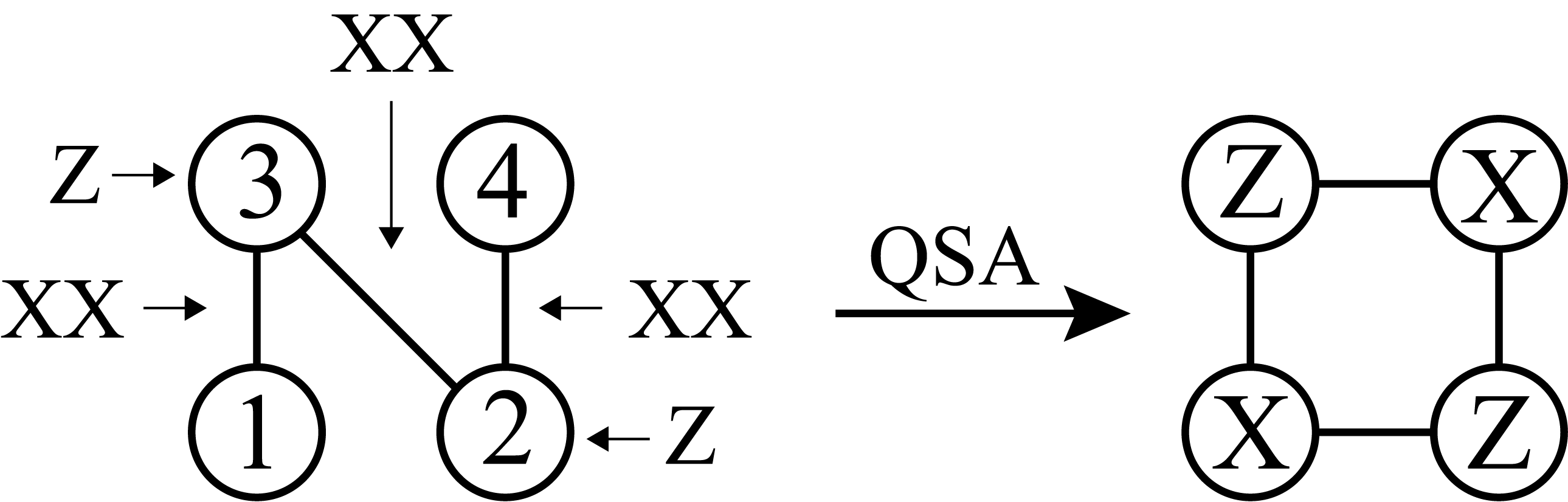}
\caption{Use QSA to simulate the four-body interaction $X_1 Z_2 Z_3 X_4$ from controlled Hamiltonians $X_1 X_3+Z_3, \, X_2 X_4+Z_2,$ and $X_2 X_3$.}
\label{fig:toric QSA}
\end{figure}

The setup of controlled and target Hamiltonians is shown in Fig. \ref{fig:toric QSA}, which is intentionally similar to the past paper \cite{hybrid4} so one could directly use the digital quantum simulation in the past paper \cite{hybrid4} after QSA without any modifications. The target Hamiltonian is $X_1 Z_2 Z_3 X_4$, which is one of the individual plaquette operators $P_{i, j}$. Other plaquette operators are the same with this target Hamiltonian but on different spins. The controlled Hamiltonians are $X_1 X_3+Z_3, \, X_2 X_4+Z_2,$ and $X_2 X_3$, which can be thought as three $XX$ two-body interactions and two $Z$ single-body Hamiltonians. All two-body interactions are chosen to be the same type to decrease the difficulty of experimental realization.

The QSA procedure here is straightforward, which is simply a particular example of the four-body QSA shown in Eq. \eqref{eq:QSA general four}. The original Hamiltonian is $X_2 X_3$, and the attachment Hamiltonians are $X_1 X_3+Z_3$ and $X_2 X_4+Z_2$. The QSA processes for those two attachment Hamiltonians are done simultaneously, given by
\be
\begin{aligned}
& e^{-i \Omega_2 \tau_2\left(X_2 X_4+Z_2\right)} e^{-i \Omega_3 \tau_3\left(X_1 X_3+Z_3\right)} e^{-i g_{23} \tau_{\alpha} X_2 X_3} \\
& \times e^{-i \Omega_3^{\prime} \tau_3^{\prime}\left(X_1 X_3+Z_3\right)}e^{-i \Omega_2^{\prime} \tau_2^{\prime}\left(X_2 X_4+Z_2\right)}\\
=&e^{-i J \tau X_1 Z_2 Z_3 X_4} 
\end{aligned}
\label{eq:toric QSA four}
\ee
where $\Omega_i \tau_i=-\frac{ \pi}{2}, \, \Omega_i^{\prime} \tau_i^{\prime}=\frac{ \pi}{2}$, and $ J \tau = g_{23} \tau_{\alpha}$. As a side note, in the rest of this paper, we will omit the illustration of how to construct any specific QSA, like Eq. \eqref{eq:toric QSA four}, to avoid excessively repeating the same discussion, because those are just particular examples of the general QSA demonstrated in Sec. \ref{sec:Quantum Simulation by Attachment}.

Now, we have simulated individual plaquette operators $P_{i, j}$ using QSA. The next step is to use digital quantum simulation to simulate the whole toric code Hamiltonian $H_w$ from individual plaquette operators $P_{i, j}$. This digital quantum simulation step is exactly the same as the digital step in the past paper \cite{hybrid4} due to the intentionally designed target Hamiltonian of four-body QSA. This digital step will be briefly summarized in the following.

\begin{figure}[t]
\centering
\includegraphics[width=\columnwidth]{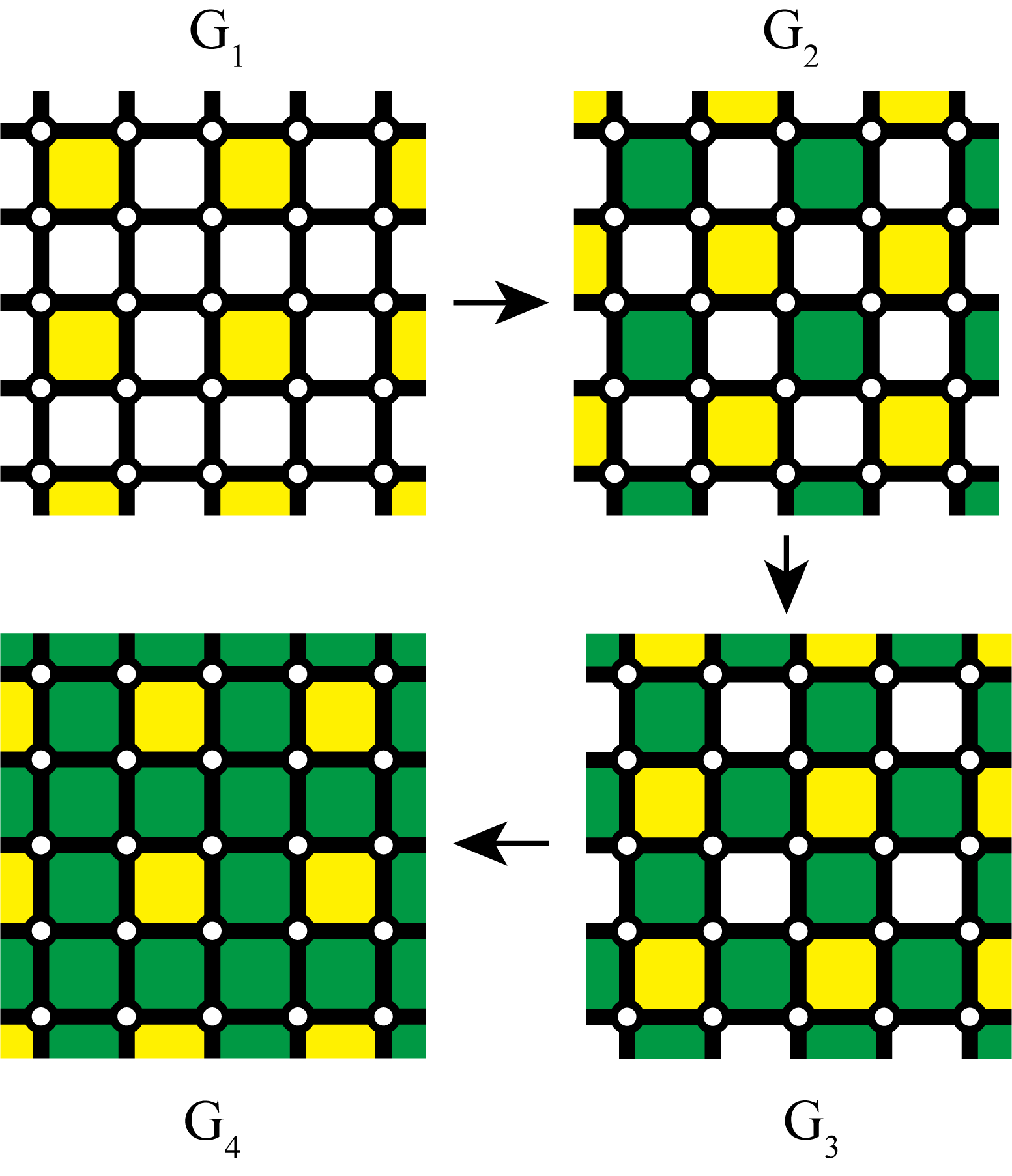}
\caption{Digital quantum simulation for simulating the whole toric code Hamiltonian $H_w$ from individual plaquette operators $P_{i, j}$. In each step, a group $G_\alpha$ of plaquettes is simulated by QSA. The whole lattice is distinguished into four groups and thus is simulated after four steps. In a particular step, the group of plaquettes currently under QSA driven is labelled yellow, and the group of plaquettes simulated already is labelled green.}
\label{fig:toric digital}
\end{figure}

The central idea of the digital step for simulating the toric code is
\be
e^{-i \tau H_w}=e^{i J \tau \sum_{i, j} P_{i, j}}=\prod_{i,j} e^{i J \tau P_{i, j}},
\label{eq:toric QSA digital target}
\ee
which is valid because all $P_{i, j}$ commute with each other as $\left[P_{i, j}, P_{k, l}\right]=0$ for all possible i, j, k, l. Furthermore, one can divide $H_w$ into four groups $G_\alpha$ of plaquette operators $P_{i, j}$, where each group is composed of $P_{i, j}$ that do not act on the same spin. All $P_{i, j}$ in the same group can be simulated by QSA simultaneously since those $P_{i, j}$ are independent of each other, which is one step of the digital quantum simulation. As shown in Fig. \ref{fig:toric digital}, the whole toric code lattice can thus be simulated in four steps by digital quantum simulation. It can also be represented as
\be
\prod_{i,j} e^{i J \tau P_{i, j}}=\prod_{\alpha=1}^4 \prod_{(i, j) \in G_\alpha} e^{i J \tau P_{i, j}}.
\label{eq:toric QSA digital target grouped}
\ee
In the digital quantum simulation, we only need propagators $\exp (i J \tau P_{i, j})$ of $P_{i, j}$, where those propagators are exactly the simulation results of four-body QSA. It demonstrates that we indeed can simulate the toric code Hamiltonian using this modified hybrid quantum simulation where the analog quantum simulation step is replaced by the QSA.

\section{Ground State and Anyons of the Toric Code}
\label{sec:Ground State and Anyons of the Toric Code}

Topological quantum computation is based on quasiparticles known as anyons \cite{RMPreviewTQC,Pbook}. Unlike bosons and fermions, exchanging two anyons can generate a phase different than $\pm 1$ or even a unitary evolution. Anyons that can generate a unitary evolution are known as non-Abelian anyons, whereas others are known as Abelian anyons. The braiding of non-Abelian anyons, which moves anyons around each other, can generate unitary evolutions that are then employed to perform quantum computation. It is the basic mechanism of topological quantum computation.

As a common model of topological quantum computation, the toric code follows the same principle of using anyons to perform quantum computation. In particular, in the toric code, the anyons are generated and braided by applying Pauli strings to the ground state of the toric code Hamiltonian. Anyons in the toric code are Abelian anyons, whereas in modified toric codes, anyons are non-Abelian. The method of generating and braiding anyons in modified toric codes is similar to the toric code, with some straightforward modifications. In this section, we will introduce the application of QSA on generating the toric code ground state as well as generating and braiding anyons on this toric code ground state. Before we begin, it is worth highlighting that we are constructing instead of simulating the ground state and anyons in the toric code, which will be further explained later in this section. It means the procedure introduced here for realizing the ground state and anyons is valid for a real toric code Hamiltonian as well as for a simulated toric code Hamiltonian in Sec \ref{sec:Simulating the Toric Code Hamiltonian}.

\subsection{Ground State of the Toric Code}

\begin{figure}[t]
\centering
\includegraphics[width=\columnwidth]{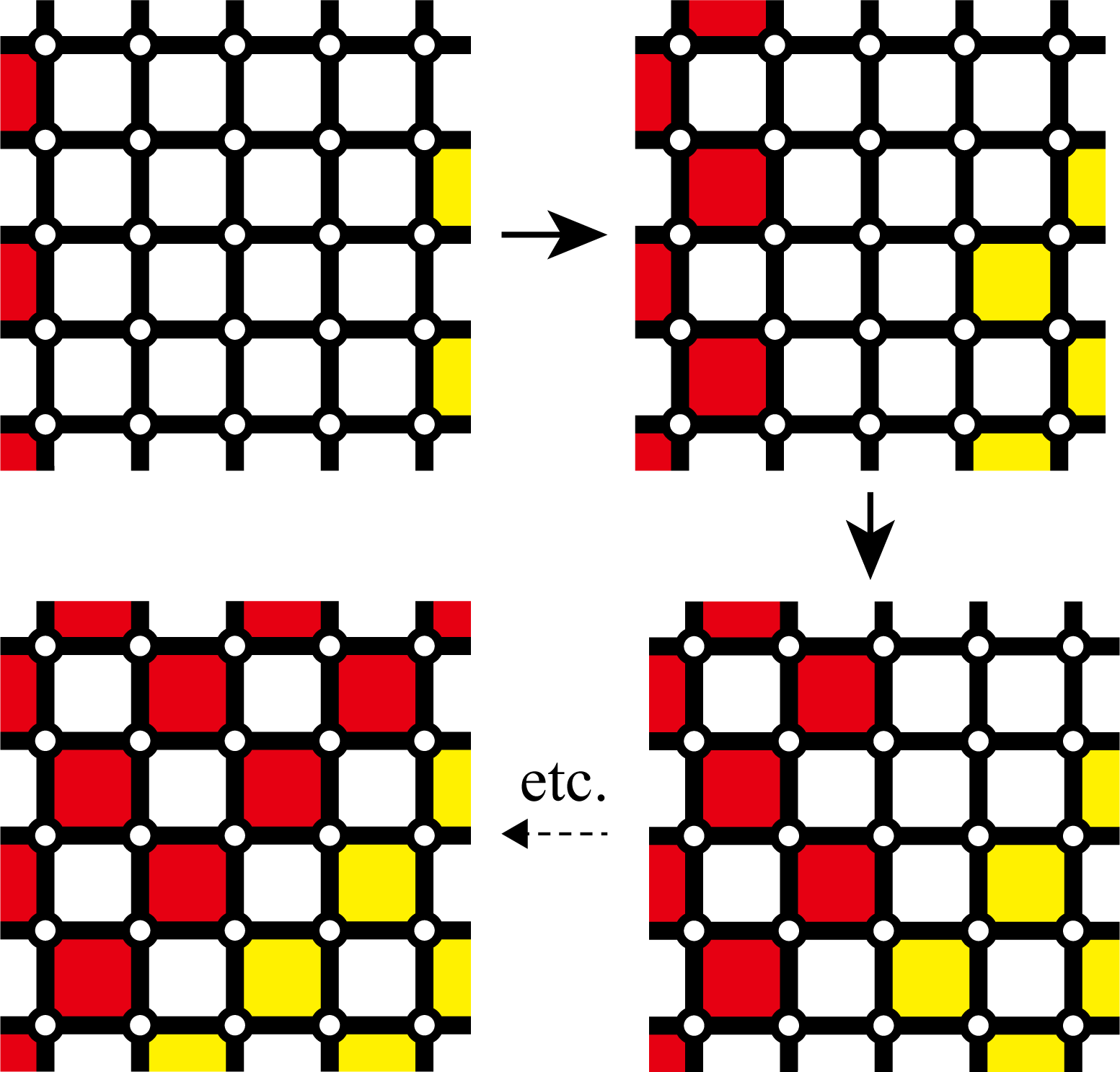}
\caption{Generating the ground state of the toric code by orderly applying $U_{i, j}^A$ and $U_{i, j}^B$ on $\left|\psi_0\right\rangle$. The red color indicates $U_{i, j}^A$, and the yellow color indicates $U_{i, j}^B$. $U_{i, j}^A$ and $U_{i, j}^B$ are acting on diagonals where $i+j$ is even. In each time step, $U_{i, j}^A$ and $U_{i, j}^B$ increase one plaquette along each diagonal. It takes six steps to reach the last figure, where some mediate steps are omitted.}
\label{fig:anyon ground}
\end{figure}

In order to construct the ground state of the toric code, we follow a similar method as the paper \cite{hybrid4}, which will be briefly introduced now. The ground state $|G\rangle$ is given as
\be
\begin{aligned}
& |G\rangle=\prod_{i+j \text { even }} \frac{\left(1+P_{i, j}\right)}{\sqrt{2}}\left|\psi_0\right\rangle, \\
& \left|\psi_0\right\rangle=\prod_{i+j \text { odd }}\frac{\left(X_{i, j}+Z_{i, j}\right)}{\sqrt{2}}|0\rangle^{\otimes N}
\end{aligned}
\label{eq:anyon ground def}
\ee
on the lattice shown in Fig. \ref{fig:toric wen} where the bottom-left spin is chosen to have index $(1,1)$. $\left|\psi_0\right\rangle$ can be directly realized, but the operator $\left(1+P_{i, j}\right)$ cannot be directly implemented since it is not unitary. However, one can generate it in another manner. We define
\be
\begin{aligned}
U_{i, j}^A& =\exp \left(-i \pi A_{i, j} / 4\right) \\
&=\exp \left(-i \pi X_{i, j} Z_{i, j+1} Z_{i+1, j} Y_{i+1, j+1} / 4\right), \\
U_{i, j}^B&=\exp \left(-i \pi B_{i, j} / 4\right)\\
&=\exp \left(-i \pi Y_{i, j} Z_{i, j+1} Z_{i+1, j} X_{i+1, j+1} / 4\right),
\end{aligned}
\label{eq:anyon ground evolution}
\ee
where $U_{i, j}^A$ and $U_{i, j}^B$ are acting on plaquettes with $i+j$ being even. One can obtain the ground state $|G\rangle$ by acting $U_{i, j}^A$ and $U_{i, j}^B$ on state $\left|\psi_0\right\rangle$ following a proper order, which is shown in Fig. \ref{fig:anyon ground}. $U_{i, j}^A$ and $U_{i, j}^B$ are applied on $\left|\psi_0\right\rangle$ in diagonal direction, where $U_{i, j}^A$ are from bottom-left to top-right and $ U_{i, j}^B$ are from top-right to bottom-left. $U_{i, j}^A$ starts from left boundary and $U_{i, j}^B$ starts from right boundary. $U_{i, j}^A$ and $U_{i, j}^B$ increase one plaquette on each diagonal, which can be treated as taking one time step. Thus, the total time required for generating the ground state increases linearly if the length of the longest diagonal in the toric code lattice increases linearly.

Everything till now about generating the ground state has already been described in the past paper \cite{hybrid4}. We are now ready to impose QSA for generating $U_{i, j}^A$ and $U_{i, j}^B$. Instead of using
\be
\begin{aligned}
& U_{i, j}^A=e^{i \frac{\pi}{4} Z_{i+1, j+1}} e^{i \frac{\pi}{4} P_{i, j}} e^{-i \frac{\pi}{4} Z_{i+1, j+1}},\\
& U_{i, j}^B=e^{i \frac{\pi}{4} Z_{i, j}} e^{i \frac{\pi}{4} P_{i, j}} e^{-i \frac{\pi}{4} Z_{i, j}} 
\end{aligned}
\label{eq:anyon ground evolution generate original}
\ee
in the past paper \cite{hybrid4}, we can directly use four-body QSA to generate $U_{i, j}^A$ and $U_{i, j}^B$, which simplifies the procedure by reducing two propagators before and after $ e^{i \frac{\pi}{4} P_{i, j}}$ in Eq. \eqref{eq:anyon ground evolution generate original}.

It is important to note that the fundamental mechanism of QSA is to simulate a target Hamiltonian by generating a propagator from controlled Hamiltonians that is the same as the propagator of the target Hamiltonian. It means propagators themselves are identical between controlled and target Hamiltonian; that is not a simulation, but it is just the same. Therefore, the ground state generated here using QSA is exactly the required ground state instead of a simulation of the required ground state since the ground state is entirely generated by propagators as in Eq. \eqref{eq:anyon ground def} where the propagator obtained by QSA is precisely the required propagator instead of a simulation. The same property will also be found in the case of generating anyons discussed later, where anyons are generated instead of simulated by QSA.

\subsection{Anyons in the Toric Code}

Anyons play a central role in topological quantum computation. The protection of quantum information against errors in the toric code depends on the size of the toric code lattice, which makes generating and braiding anyons in a large lattice practically important \cite{Pbook}. In order to generate and braid anyons in a large lattice, one needs to apply long Pauli strings $ \sigma_1 \sigma_2 \ldots \sigma_N $ or their propagators $\exp (-i t g \sigma_1 \sigma_2 \ldots \sigma_N)$ on the spins of the lattice. Although in some topological quantum computation models, like the original toric code \cite{Ktoric} and the toric code with twists \cite{twist}, generating and braiding anyons need only long Pauli strings $\sigma_1 \sigma_2 \ldots \sigma_N$, but in other models, like the toric code quantum memory \cite{Pbook} and the toric code with holes \cite{hole}, propagators $\exp (-i t g \sigma_1 \sigma_2 \ldots \sigma_N)$ is required for generating and braiding anyons.

Long Pauli strings $\sigma_1 \sigma_2 \ldots \sigma_N$ do not cause a problem for experimental realization since acting a whole long Pauli string $\sigma_1 \sigma_2 \ldots \sigma_N$ on the lattice is the same with separately acting each Pauli operator $\sigma_i$ of the string on the corresponding spins of the lattice. However, propagators $\exp (-I t g \sigma_1 \sigma_2 \ldots \sigma_N)$ cannot be realized in this simple manner. Realizing long Pauli propagators using two-body interactions between neighboring spins remains a challenge \cite{hybrid4,nbodyS}. The ability to overcome this challenge is a significant advantage of using QSA in topological quantum computation. In this subsection, we will first introduce anyons in the toric code to set up the conceptual framework for discussions on the toric code and its modifications in the rest of this paper. After that, the application of QSA on generating and braiding those anyons is demonstrated. The quantum memory built on the toric code lattice employing those anyons is then introduced as a particular example demonstrating the benefit of using QSA in the topological quantum computation, since this quantum memory requires long Pauli string propagators which cannot be generated without QSA.

\begin{figure}[t]
\centering
\subfloat[\label{fig:anyon anyon def}]{\includegraphics[width=0.45\columnwidth]{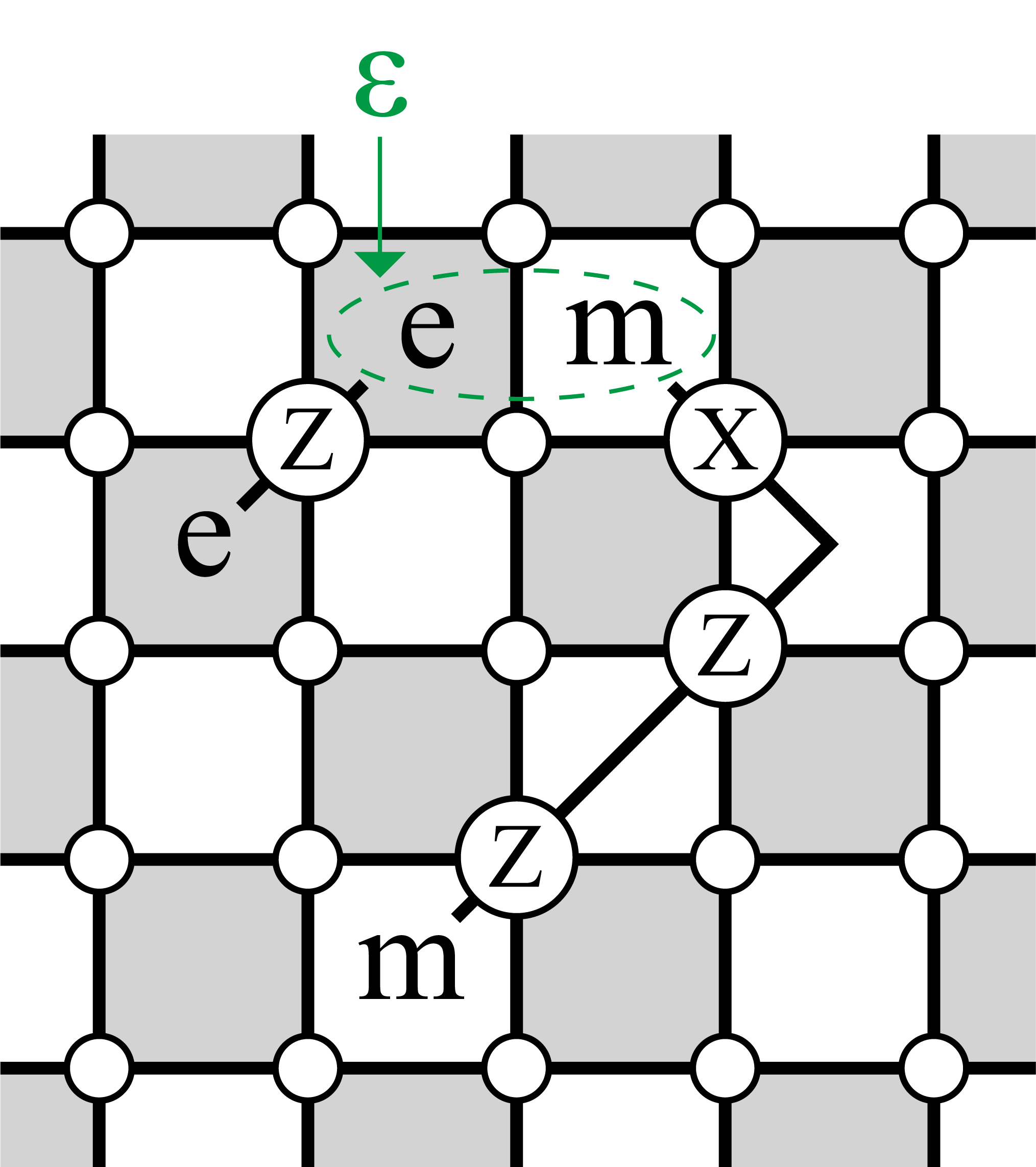}}
\hfill
\subfloat[\label{fig:anyon anyon braiding}]{\includegraphics[width=0.45\columnwidth]{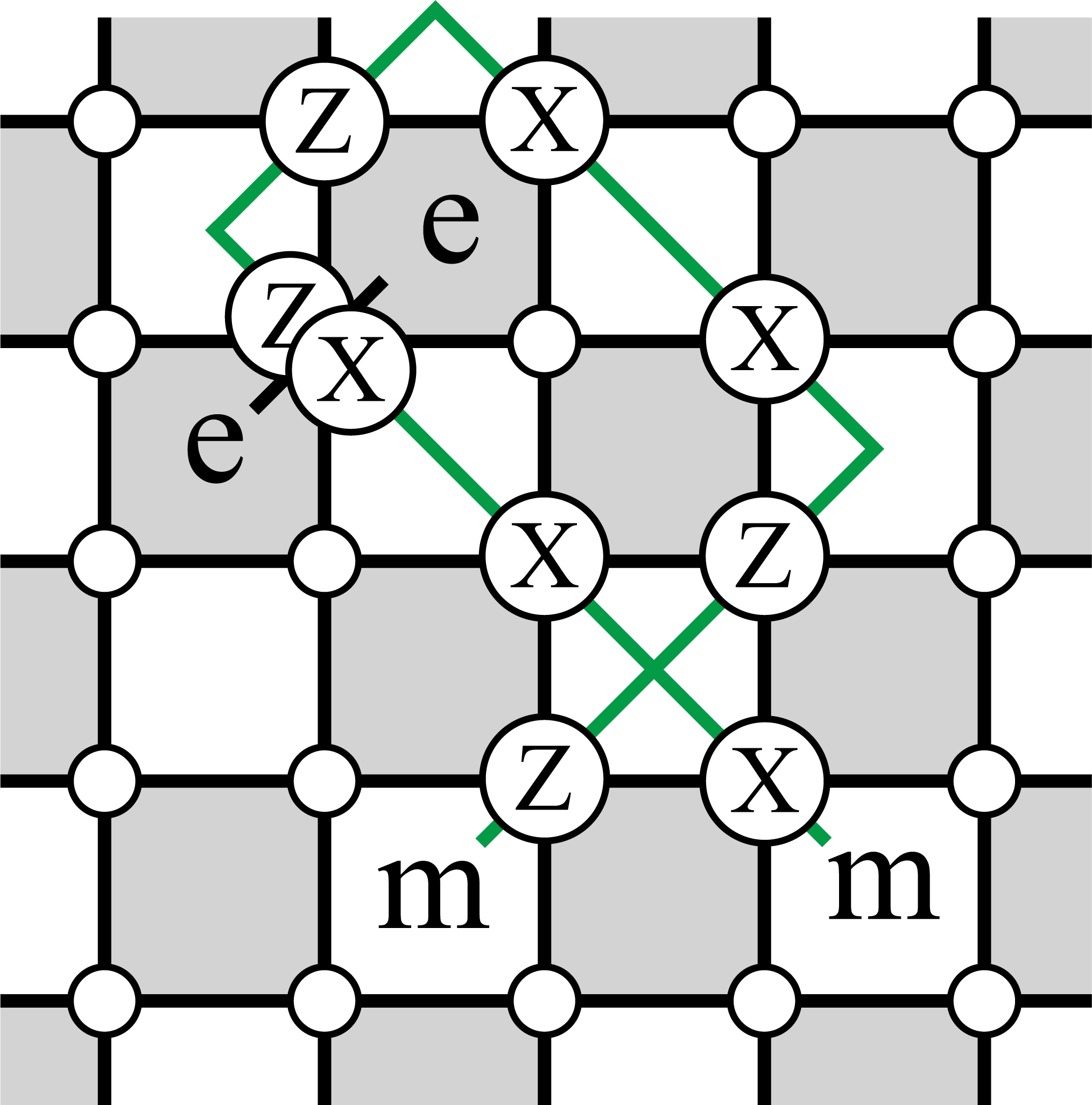}}
\label{fig:anyon anyon def and braiding}
\caption{(a) Generating and moving anyons in the toric code lattice. A pair of $e$ or $m$ anyons can be generated and moved by a Pauli string applied on spins in the lattice, where two anyons of the same kind are at the endpoints of a Pauli string. A pair of $e$ and $m$ anyons can also be treated as a whole, which is a new anyon of a different kind named as the $\epsilon$ anyon indicated by the green ellipse and arrow. (b) An $m$ anyon is braided around an $e$ anyon using a Pauli string indicated by the green line. An $X$ operator in the $m$ anyon Pauli string is applied on a spin after a $Z$ operator in the $e$ anyon Pauli string is applied on the same spin.}
\end{figure}

The toric code has three kinds of anyons, labelled as $e$, $m$, and $\epsilon$ \cite{Ktoric, Wtoric}. The anyon generating and braiding operations in the toric code are the same \cite{twist}. Both are done by applying a Pauli string on the ground state of the toric code as shown in Fig. \ref{fig:anyon anyon def}. If one applies a single Pauli $Z$ or $X$ operator on a spin in the lattice, a pair of anyons will be generated on plaquettes in the diagonal direction. Anyons are generated on bottom-left and top-right plaquettes for the Pauli $Z$ operator, and generated on top-left and bottom-right plaquettes for the Pauli $X$ operator. The lattice is distinguished into dark and light colored plaquettes, where plaquettes on the same diagonal have the same color and the two adjacent diagonals have different colors. The most bottom-left positioned plaquette is artificially chosen to be dark, although it can also be set as light since there is not any physical difference between dark and light plaquettes due to the symmetry of the toric code. Anyons generated on dark plaquettes are labelled as $e$ anyons, whereas anyons generated on light plaquettes are labelled as $m$ anyons. If two $e$ or $m$ anyons are on the same plaquette, they are annihilated with each other to become the vacuum. This means that one can use a Pauli string acting on nearby spins to generate a pair of far-apart anyons. As shown by $m$ anyons in Fig. \ref{fig:anyon anyon def}, all intermediate anyons in the Pauli string are annihilated, and only two anyons at the endpoints are reminded. This way of generating anyons using a Pauli string is equivalent to moving anyons from near each other to far apart. The braiding of anyons can also be done by using a Pauli string, as shown in Fig. \ref{fig:anyon anyon braiding}.

Pauli strings for generating and braiding anyons have the form $ \sigma_1 \sigma_2 \ldots \sigma_N$ where $\sigma_i$ is a Pauli $X$ or $Z$ operator. $\sigma_i$ operators are acting on the corresponding spins on the path of the Pauli string, as shown in Fig. \ref{fig:anyon anyon def} and Fig. \ref{fig:anyon anyon braiding}. Those Pauli strings $ \sigma_1 \sigma_2 \ldots \sigma_N$ can be directly implemented by separately applying each individual Pauli operator $\sigma_i$ on the corresponding spin, which is the usual way of implementing Pauli strings. However, as we mentioned before, we also need Pauli string propagators $\exp (-I t g \sigma_1 \sigma_2 \ldots \sigma_N)$ in certain circumstances, like the toric code quantum memory \cite{Pbook} which will be shown later in this section. Those Pauli string propagators $ \sigma_1 \sigma_2 \ldots \sigma_N$ have the spatial linear shape, so they can be directly realized using general QSA with linearly increasing spin number introduced in Sec. \ref{sec:Quantum Simulation by Attachment}. Those Pauli string propagators generated by general QSA can have an arbitrary length without requiring an experimental setup more complex than two-body interactions between next-nearest neighbors $\sigma_{i, j}\sigma_{i+1, j+1}$.

The main reason why Pauli string propagators $ \sigma_1 \sigma_2 \ldots \sigma_N$ are essential in certain circumstances is that Pauli string propagators can be rewritten as
\be
e^{-i t g \sigma_1 \sigma_2 \ldots \sigma_N}=\cos (t g) \mathbb{I}-i \sin (t g) \sigma_1 \sigma_2 \ldots \sigma_N ,
\label{eq:anyon anyon evolution of braiding}
\ee
which enables us to generate a superposition of altered state and unaltered state as
\be
\begin{aligned}
&e^{-i t g \sigma_1 \sigma_2 \ldots \sigma_N}\left|\psi\right\rangle\\
=&\cos (t g) \left|\psi\right\rangle-i \sin (t g) \sigma_1 \sigma_2 \ldots \sigma_N\left|\psi\right\rangle
\end{aligned}
\label{eq:anyon anyon state after braiding}
\ee
instead of only a altered state $\sigma_1 \sigma_2 \ldots \sigma_N\left|\psi\right\rangle$ when using Pauli strings $ \sigma_1 \sigma_2 \ldots \sigma_N$. Depending on whether the Pauli string propagator is used to generate or braid anyons, this superposition of altered and unaltered states is a superposition of generated and ungenerated states or a superposition of braided and unbraided states. As a side note, one also can choose $t g$ such that $\exp (-i t g \sigma_1 \sigma_2 \ldots \sigma_N)$ degenerates into $i \sigma_1 \sigma_2 \ldots \sigma_N$ where the overall phase $i$ can be ignored, so one can recover ordinary Pauli strings $ \sigma_1 \sigma_2 \ldots \sigma_N$ in this situation. The ability to generate a superposition of altered and unaltered states introduces a wider range of possible resulting states in one operation, which is essential in many circumstances. As an example of the importance of this ability, we will now introduce the case of employing the toric code to build a topologically protected quantum memory \cite{Pbook}.

\begin{figure}[t]
\centering
\includegraphics[width=\columnwidth]{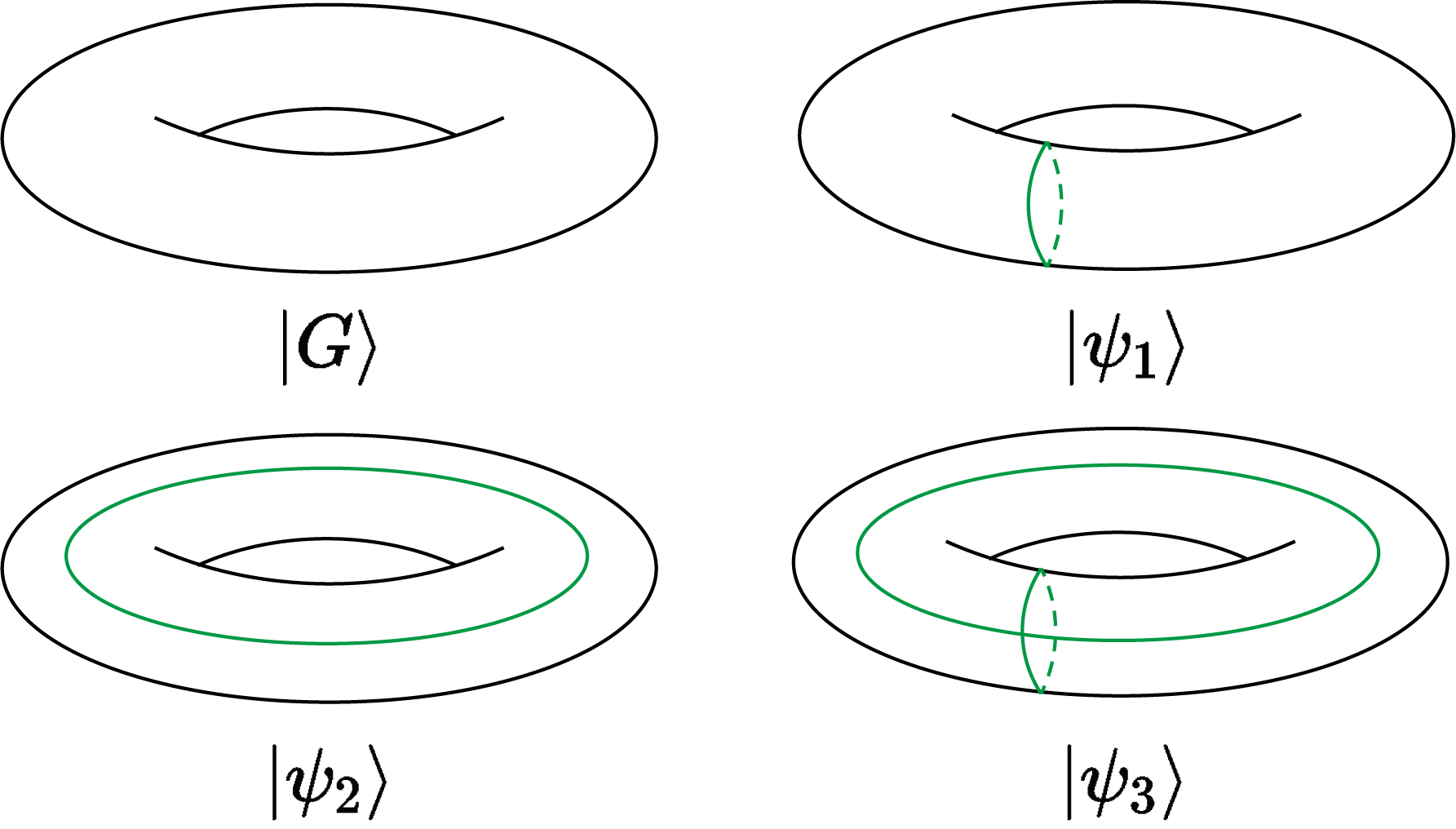}
\caption{Four states for storing quantum information in a topologically protected manner using the toric code with periodic boundary conditions. $\left|G\right\rangle$ is the ground state of the toric code. $\left|\psi_1\right\rangle$, $\left|\psi_2\right\rangle$ and $\left|\psi_3\right\rangle$ can be obtained by applying Pauli strings which form a vertical loop, a horizontal loop, and both vertical and horizontal loops correspondingly, where green lines are Pauli strings. Quantum information in this toric code quantum memory is stored as a superposition of $\left|G\right\rangle$, $\left|\psi_1\right\rangle$, $\left|\psi_2\right\rangle$, and $\left|\psi_3\right\rangle$.}
\label{fig:anyon anyon toric memory}
\end{figure}

As we mentioned at the beginning of this section, anyons in the toric code are all Abelian anyons that cannot support quantum computation. However, the toric code can work as a topologically protected quantum memory as shown in the past research \cite{Pbook}, which will be briefly introduced here. In order to build the quantum memory, the toric code lattice is set to have periodic boundary conditions. As shown in Fig. \ref{fig:anyon anyon toric memory}, one can vertically or horizontally apply a Pauli string on this lattice to form a loop, where anyons at the endpoints of the Pauli string will annihilate with each other. We artificially choose $e$ anyons to form loops, although $m$ anyons is also a valid choice. There are four different states $\left|G\right\rangle$, $\left|\psi_1\right\rangle$, $\left|\psi_2\right\rangle$, and $\left|\psi_3\right\rangle$ obtainable from those Pauli loops, which are corresponding from no loop, a vertical loop, a horizontal loop, and both vertical and horizontal loops. Those states are topologically protected, since an error must also be generated by a Pauli string loop which is unlikely to naturally occur. Those states are thus employed to store quantum information in the toric code quantum memory. Non-trivial quantum information usually needs to be represented as a superposition of quantum states. It means that individual states of $\left|G\right\rangle$, $\left|\psi_1\right\rangle$, $\left|\psi_2\right\rangle$, and $\left|\psi_3\right\rangle$ generated by Pauli string operators $\sigma_1 \sigma_2 \ldots \sigma_N$, are not sufficient to build a meaningful quantum memory, since one cannot build a superposition of those states using Pauli string operators. However, the superposition of those states can be generated by Pauli string propagators $ \sigma_1 \sigma_2 \ldots \sigma_N$, where the propagators are produced by the general QSA. The superposition of states is also topologically protected, because an error must be generated by a Pauli string loop propagator which is unlikely to naturally occur. The ability to generate and store the superposition of states enables the toric code to function as a topologically protected quantum memory. As a side note, in Eq. (\ref{eq:anyon anyon state after braiding}), one may notice that we are not able to generate an arbitrary superposition of two states using only Pauli loop propagators that generate $e$ anyons. This issue can be overcome by also using Pauli loop propagators which generate $m$ anyons \cite{DennisQuantumMemory,hole} along with Pauli loop propagators that generate $e$ anyons, which will be introduced in the discussion about the toric code with holes \cite{hole} in Sec. \ref{sec:Simulating Modified Toric Codes}.

\section{Simulating Modified Toric Codes}
\label{sec:Simulating Modified Toric Codes}

In Sec. \ref{sec:Simulating the Toric Code Hamiltonian} and Sec. \ref{sec:Ground State and Anyons of the Toric Code}, we have shown how to use QSA to simulate the toric code Hamiltonian as well as to generate and braid anyons for performing topological quantum computation on the toric code. In this paper, we will also work on two modified toric codes, which are the toric code with twists \cite{twist} and the toric code with holes \cite{hole}. Both modified toric codes can support universal quantum computation directly or indirectly \cite{twist,hole}, whereas the original toric code \cite{Ktoric} cannot do so. Using QSA to simulate Hamiltonians and to generate and braid anyons in modified toric code is mostly the same as doing that in the toric code. We can use the hybrid quantum simulation \cite{hybrid3,hybrid4} and QSA to simulate Hamiltonians in modified toric codes. If Pauli strings alone are not sufficient for generating and braiding anyons in modified toric codes, we can use QSA to obtain Pauli string propagators to do the task. In this section, we will focus on the major differences between modified toric codes and the toric code. We will demonstrate how to modify the approaches used in the original toric code such that those approaches can be employed in modified toric codes to simulate Hamiltonians as well as to generate and braid anyons.

\subsection{Toric Code with Twists}

\begin{figure}[t]
\centering
\includegraphics[width=0.75\columnwidth]{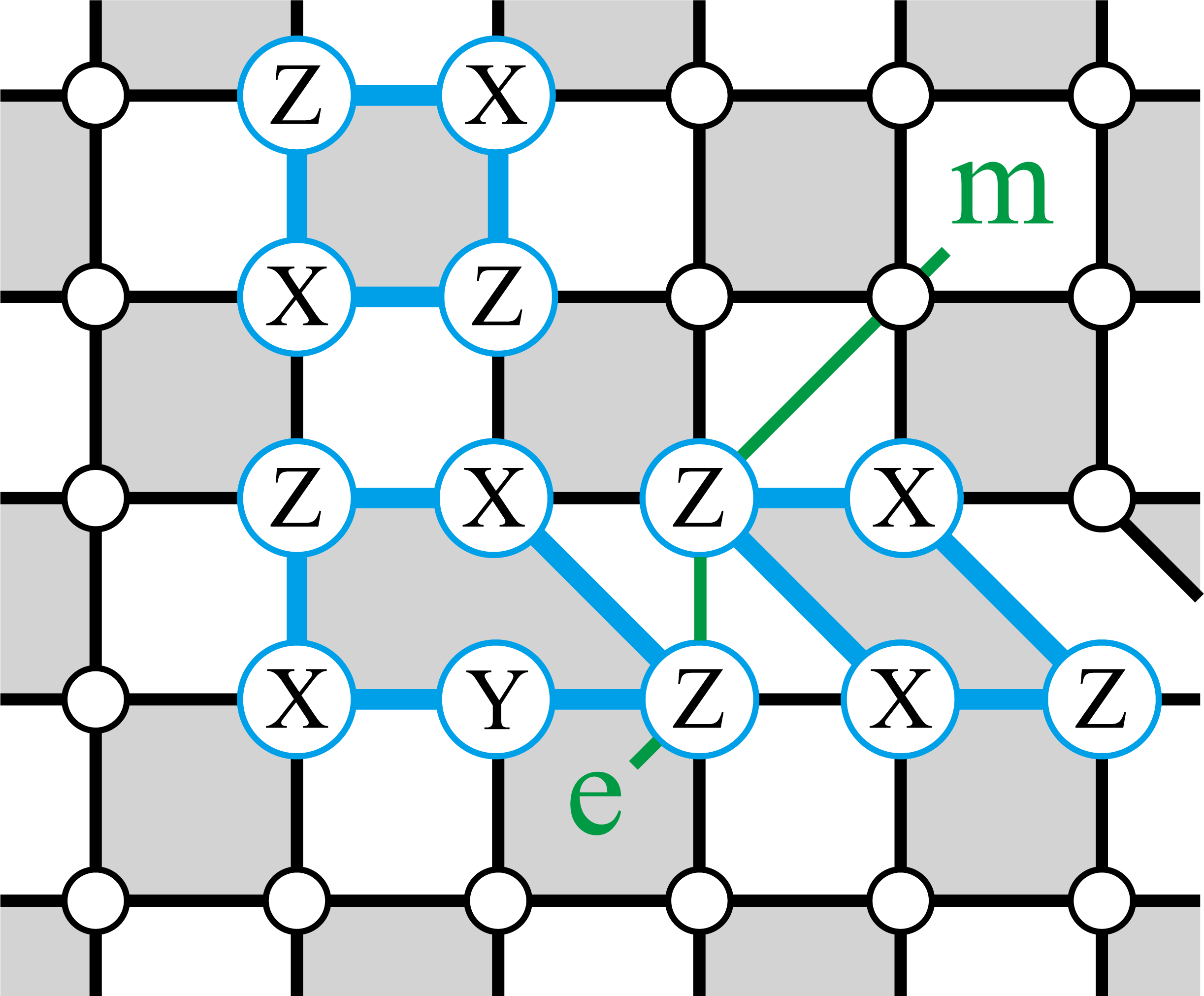}
\caption{Toric code with twists. There are five-body interactions $XZYZX$, known as twists, inserted in the original toric code lattice composed of four-body interactions $XZZX$. Four-body interactions between a pair of twists are deformed without changing their Hamiltonian. As indicated by the green line, an $e$ anyon that goes across those deformed four-body interactions changes its kind to be an $m$ anyon.}
\label{fig:modified twist}
\end{figure}

The main difference between the toric code with twists \cite{twist} and the original toric code \cite{Wtoric} in Sec. \ref{sec:Simulating the Toric Code Hamiltonian} is their Hamiltonians. Excepting the original four-body interactions $XZZX$, there are also five-body interactions of form $XZYZX$ in the Hamiltonian of the toric code with twists as shown in Fig. \ref{fig:modified twist}. Those five-body interactions are twists, and four-body interactions between a pair of twists are spatially deformed. Anyons that go across those deformed four-body interactions will change their kind, where $e$ anyons will become $m$ anyons and vice versa. This change of kind enables the toric code with twists to perform quantum computation \cite{twist}. However, one still needs some other topologically unprotected operations to make the quantum computation in the toric code with twists being universal \cite{isinguniversal}, which is out of the scope of this paper. The five-body interactions $XZYZX$ are the only relevant difference in toric code with twists since deformed four-body interactions do not affect how to use QSA to simulate the Hamiltonian and to generate and braid anyons. The rest of this subsection is thus mainly about those five-body interactions.

Simulating the toric code with twists using QSA is straightforward. Five-body interactions $XZYZX$ commute with nearby four-body interactions and surly also commute with distanced four-body and five-body interactions. It indicates that the framework of hybrid quantum simulation \cite{hybrid3,hybrid4} used in Sec. \ref{sec:Simulating the Toric Code Hamiltonian} for simulating the original toric code \cite{hybrid4} is also valid for simulating the toric code with twists. The only modification is on simulating individual Pauli strings, where five-body interactions $XZYZX$ also need to be simulated besides four-body interactions. However, it is not an issue for QSA since simulating five-body interactions is just a particular example of the general QSA for five-body interactions. Five-body interactions $XZYZX$ can thus be directly simulated by this five-body QSA, which enables the Hamiltonian of the toric code with twists being simulated by QSA. Generating and braiding anyons in the toric code with twists can be done in the same manner as the case of the original toric code in Sec. \ref{sec:Ground State and Anyons of the Toric Code}.

\subsection{Toric Code with Holes}

\begin{figure}[t]
\centering
\subfloat[\label{fig:modified holes Hamiltonian and anyons}]{\includegraphics[width=0.45\columnwidth]{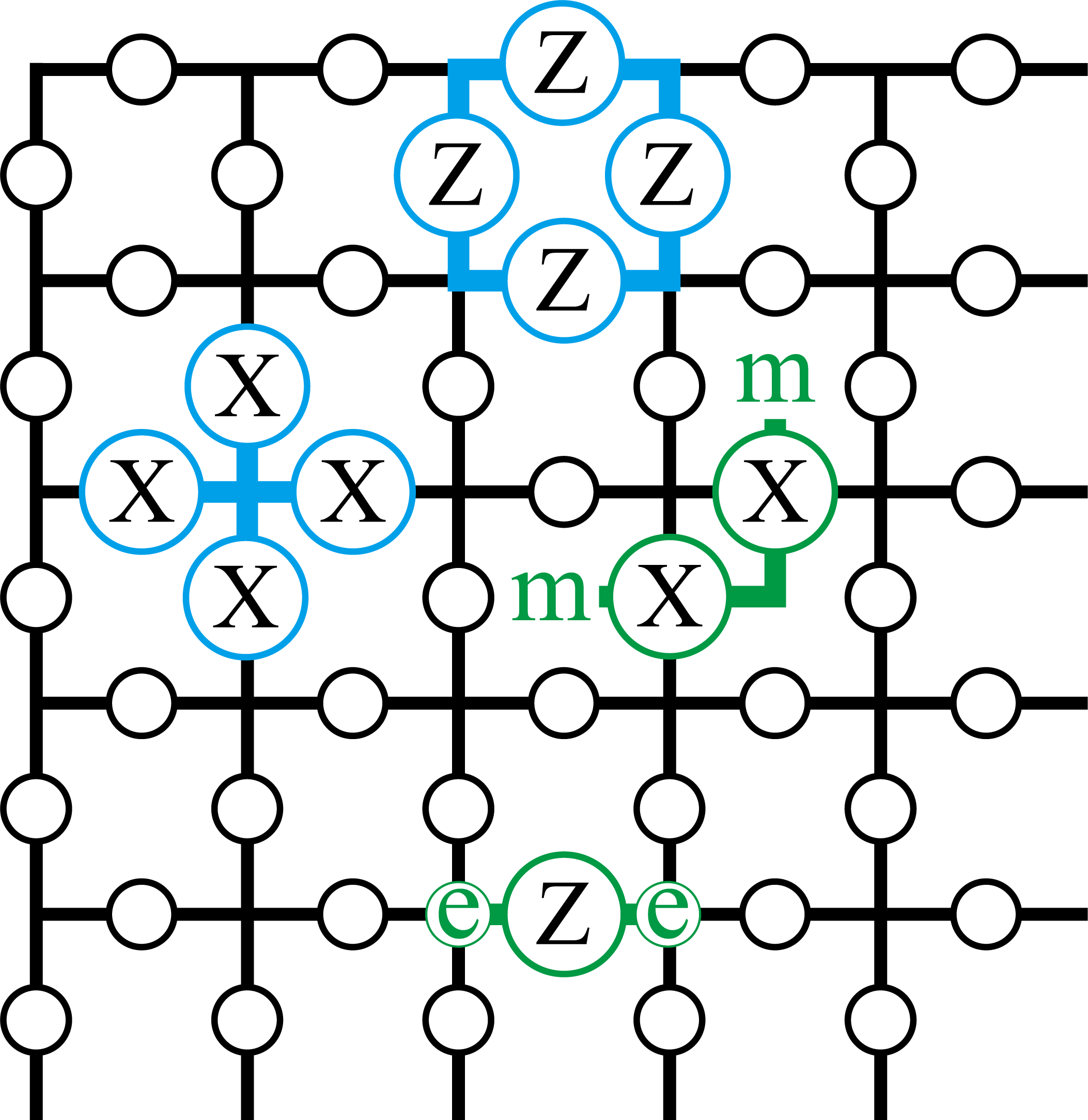}}
\hfill
\subfloat[\label{fig:modified holes holes}]{\includegraphics[width=0.515\columnwidth]{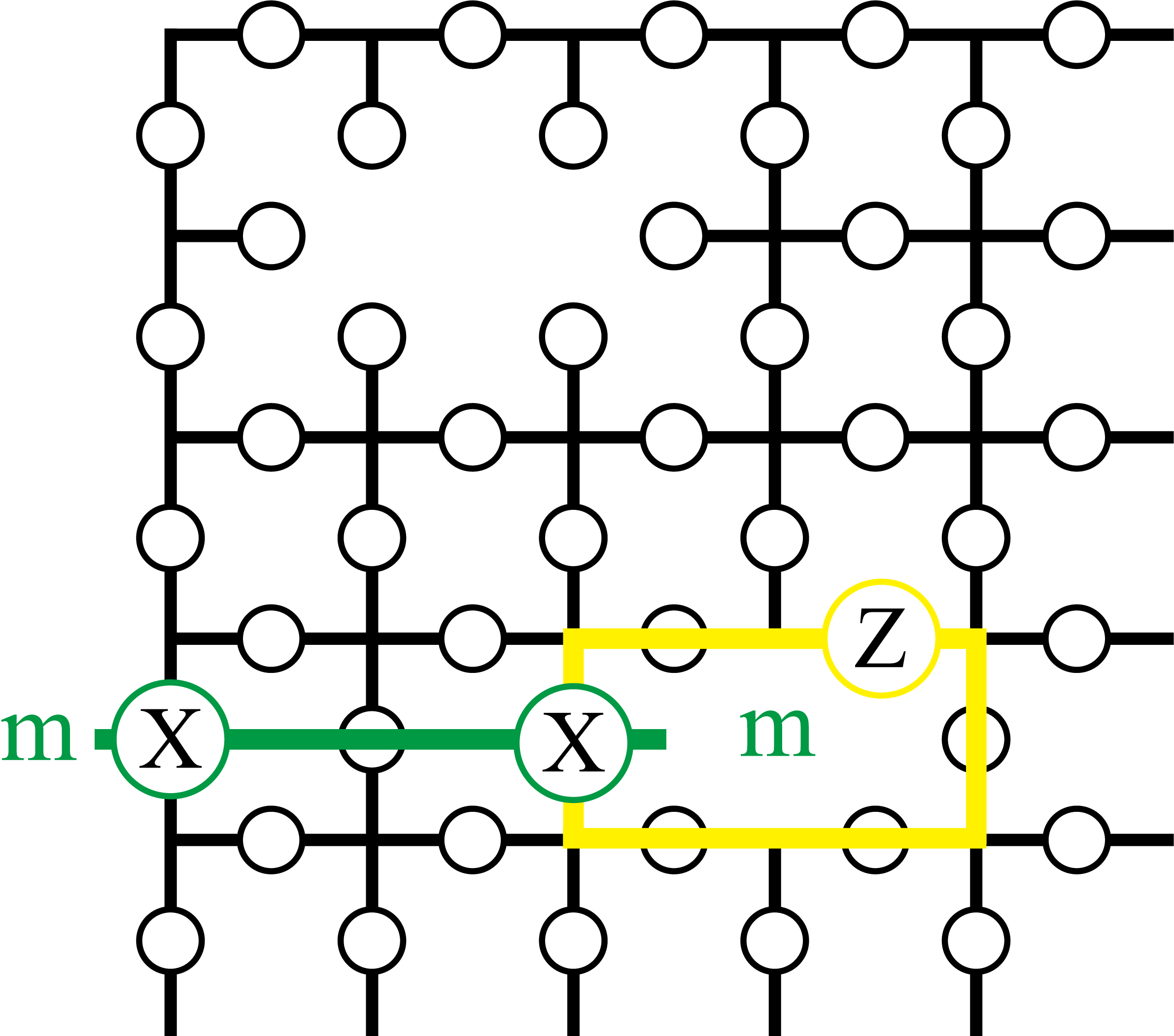}}
\label{fig:modified holes group fig}
\caption{(a) Lattice of the toric code with holes, where holes are not shown in this figure. The Hamiltonian is composed of $ZZZZ$ on all plaquettes and $XXXX$ on all vertices, as indicated by blue lines. Pauli strings of only $X$ operators can generate $m$ anyons on plaquettes, and Pauli strings of only $Z$ operators can generate $e$ anyons on vertices, as shown by green lines. (b) Two holes on the lattice, where the hole on the bottom right is a smooth hole and the hole on the top left is a rough hole. }
\end{figure}

There are two major differences between the toric code with holes \cite{hole} and the original toric code. Firstly, instead of $XZZX$ used in Sec. \ref{sec:Simulating the Toric Code Hamiltonian}, the four-body interactions in the toric code with holes are $ZZZZ$ acting on all plaquettes and $XXXX$ acting on all vertices as shown in Fig. \ref{fig:modified holes Hamiltonian and anyons}, which in fact is another equivalent form of the original toric code Hamiltonian \cite{Ktoric}. The consequence of this change is that, as shown in Fig. \ref{fig:modified holes Hamiltonian and anyons}, the graphical structure of the lattice and the spatial place of anyons in the toric code with holes are different from the original toric code. Spins in the lattice are on the sides of plaquettes instead of vertices, which does not produce any physical difference since plaquettes are only for demonstration purposes instead of being physical objects. $m$ anyons can be generated and braided on plaquettes using Pauli strings composed by Pauli $X$ operators alone, and $e$ anyons can be generated and braided on vertices using Pauli strings composed by Pauli $Z$ operators alone.

Secondly, as the name suggests, there are holes in the lattice of this toric code with holes \cite{hole} as shown in Fig. \ref{fig:modified holes holes}. There are two kinds of holes, known as smooth holes and rough holes, which can correspondingly contain $m$ anyons and $e$ anyons. Those holes, including anyons in the holes, should be able to move. Similar to holes, there are two kinds of boundaries in the lattice, which are smooth and rough boundaries, instead of periodic boundaries used in the original toric code quantum memory \cite{Pbook} in Sec. \ref{sec:Ground State and Anyons of the Toric Code}. One can use a Pauli $X$ string to connect a smooth boundary and a smooth hole to store an $m$ anyon in the hole and throw another $m$ anyon out of the lattice. If one treats a smooth hole with an $m$ anyon as $|1\rangle$ state and without an $m$ anyon as $|0\rangle$, this Pauli $X$ string is a logical $X$ operation. The logical $Z$ operation is to use Pauli $Z$ string to form a loop around a smooth hole. The rough holes with and without an $e$ anyon are treated as $|-\rangle$ and $|+\rangle$ states correspondingly. A Pauli $Z$ string connecting a rough hole and a rough boundary is a logical $Z$ operation, and a Pauli $X$ string loop around a rough hole is a logical $X$ operation. The information of those qubits is better protected if holes are larger and further away from the boundary.

In order to perform universal quantum computation in the toric code with holes, one also needs to realize the $CNOT$ gate and magic states \cite{magic1,magic2} besides those single qubit logical operations mentioned above \cite{hole}. The $CNOT$ gate is implemented by braiding a smooth hole around a rough hole or vice versa \cite{hole}, which is automatically realizable if one can move holes as assumed before. Magic states for smooth holes are
\be
|\theta\rangle=\frac{1}{\sqrt{2}}(|0\rangle+e^{ i \theta}|1\rangle)
\label{eq:modified holes magic states}
\ee
with $\theta$ being $ 0, \, \pi / 4, \, \pi / 2$ and $\pi$. Magic states for rough holes are the same with Eq. \eqref{eq:modified holes magic states} if we change $|0\rangle$ to $|+\rangle$ and change $|1\rangle$ to $|-\rangle$. 

As a summary of all information about the toric code with holes introduced above, one will be able to simulate the toric code with holes and to perform universal quantum computation on it if one can handle those two major differences between the toric code with holes and the original toric code as well as generate magic states on this model. It is a challenging task since long Pauli string propagators are required here as we will show soon, but it can be solved by QSA. Simulating the Hamiltonian and holes of the toric code with holes is straightforward, which is a modification of the approach for simulating the toric code with holes introduced in the past paper \cite{hybrid4} where QSA is employed now to replace the four-body interaction simulation used in the past paper \cite{hybrid4}. Four-body interactions in the Hamiltonian of the toric code with holes commute with each other, so one can use hybrid quantum simulation \cite{hybrid4} and QSA to simulate this Hamiltonian in a similar way with the case of original toric code in Sec. \ref{sec:Simulating the Toric Code Hamiltonian}. Simulating holes on the lattice can be done by not driving corresponding QSA in the position of those holes during the whole Hamiltonian simulation procedure.

Generating magic states and moving holes are trickier. Magic states in Eq. \eqref{eq:modified holes magic states} as superpositions of states can be generated by Pauli string propagators connecting a hole and a boundary, which is similar to the storing quantum information via Pauli string loop propagators in the toric code quantum memory \cite{Pbook} introduced in Sec. \ref{sec:Ground State and Anyons of the Toric Code}. Taking the superposition of $|0\rangle$ and $|1\rangle$ as an example, we can use general QSA to generate the Pauli string propagator corresponding to the logical $X$ propagator, which can produce a state of form
\be
\begin{aligned}
&e^{-i t g X_{\text{logical}}}\left|0\right\rangle\\
=&\cos (t g) \left|0\right\rangle-i \sin (t g) \left|1\right\rangle .
\end{aligned}
\label{eq:modified holes logical X result}
\ee
It is not a general superposition state yet since the phase difference between $|0\rangle$ and $|1\rangle$ is fixed. However, one can get a general superposition, including all magic states, by introducing a phase shift gate
\be
P(\theta)=\left[\begin{array}{cc}
1 & 0 \\
0 & e^{i \theta}
\end{array}\right]
\label{eq:modified holes phase shift gate}
\ee
which can change the phase of $|1\rangle$ in the superposition while leaving $|0\rangle$ unchanged. This phase shift gate $P(\theta)$ can be realized using QSA to generate the Pauli string propagator corresponding to the logical $Z$ propagator as
\be
\begin{aligned}
e^{-i t g Z_{\text{logical}}}=&\cos (t g) -i \sin (t g) Z_{\text{logical}} \\
=&\left[\begin{array}{cc}
e^{-i t g} & 0 \\
0 & e^{i t g}
\end{array}\right] .
\end{aligned}
\label{eq:modified holes phase by Z operator}
\ee
It is the required phase shift gate up to an ignorable global phase factor since
\be
e^{i t g}e^{-i t g Z_{\text{logical}}}=P(2t g) .
\label{eq:modified holes phase by Z operator global phase}
\ee
This method of combining QSA for generating the logical $X$ propagator and the logical $Z$ propagator to produce a general superposition of states can also be used in the case of the toric code quantum memory as mentioned in Sec. \ref{sec:Ground State and Anyons of the Toric Code}.

\begin{figure}[t]
\centering
\includegraphics[width=0.75\columnwidth]{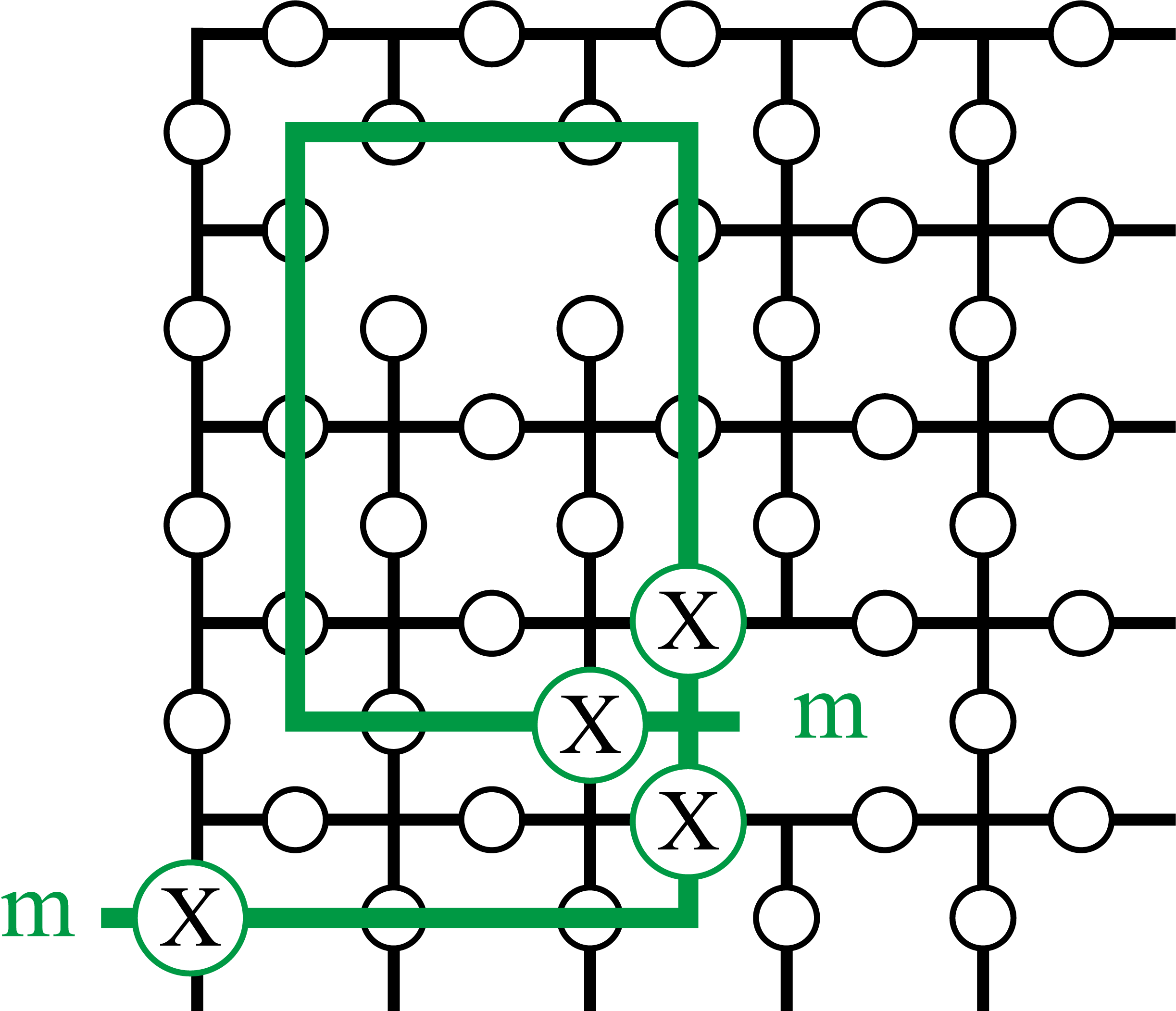}
\caption{A CNOT gate on the toric code with holes. As indicated by the green line, it is implemented by braiding a smooth hole around a rough hole for a circle using a Pauli $X$ string propagator.}
\label{fig:modified holes holes loop}
\end{figure}

We now consider how to move holes. In particular, we are interested in how to construct the CNOT gate by braiding a hole around another hole, as introduced before. Having superposition of states in holes is why moving holes is tricky. Moving an empty hole or a hole with an anyon is not difficult, as described in the past paper \cite{hybrid4}, as long as they are not superposed together. To move an empty hole, one can stop driving interactions in the new position of the moved hole and redrive interactions in the position where the moved hole left \cite{hybrid4}. For a hole with an anyon, the anyon can be moved by a single Pauli operator, and the hole can be moved in a similar way with the empty hole case \cite{hybrid4}. We can thus directly implement the CNOT gate by moving a hole around another hole if the moved hole contains an anyon or is empty. However, if the hole is in a superposition state of containing an anyon and being empty, the method of using a single Pauli operator to move the anyon in the hole will introduce some errors. A single Pauli operator acting on an anyon will move this anyon, and it acting on the empty place will generate a pair of anyons, so a single Pauli operator acting on a superposition of an anyon and empty space will introduce an error as
\be
\begin{aligned}
&X_{N+1}(\cos (t g) \left|0\right\rangle-i \sin (t g) X_1 X_2 \ldots X_N\left|0\right\rangle) \\
=&\cos (t g) X_{N+1}\left|0\right\rangle-i \sin (t g) X_1 \ldots X_NX_{N+1}\left|0\right\rangle \\
\neq &\cos (t g) \left|0\right\rangle-i \sin (t g) X_1 \ldots X_NX_{N+1}\left|0\right\rangle.
\end{aligned}
\label{eq:modified holes error by single gate}
\ee
This error is particularly harmful if the hole is close to a boundary or the hole is braided around another hole. In the former case, the Pauli string in front of $\left|0\right\rangle$ may combine with other naturally occurring Pauli $X$ errors to form a Pauli string connecting the hole and a boundary, generating a logical error. In the latter case, which is also the case of building the CNOT gate, the Pauli string in front of $\left|0\right\rangle$ forms a loop around another hole, which generates a logical error on another hole. In order to avoid this kind of error, we can directly generate the resulting superposition state of the braiding operation using the general QSA instead of braiding an existing superposition state. Taking the CNOT gate as an example, to implement the CNOT gate, we can directly use the general QSA to produce a Pauli $X$ string loop propagator around a hole, as shown in Fig. \ref{fig:modified holes holes loop}. It is identical to the result of braiding a smooth hole around a rough hole where the smooth hole is in a superposition state. However, this method of moving holes with superposition states will face certain limitations when dealing with cases more complex than the CNOT gate. Suppose there are two Pauli string propagators, A and B, crossing each other twice, where A is on the top of B at the first intersection, and B is on the top of A at the second intersection. In that case, the method of moving holes with superposition states cannot be directly employed. This is because we cannot directly use the general QSA to produce Pauli string propagator A after B nor B after A to represent this situation since Pauli string propagators A and B are mixed together in time. How to overcome this limitation is an open question for further research.

\section{Strength and Errors}
\label{sec:Strength and Errors}

We have briefly discussed the strength and errors of QSA in Sec. \ref{sec:Quantum Simulation by Attachment}. As a summary, coefficients $t g$ for the original Hamiltonian $H_O$ and $t^{\prime} g^{\prime}$ for the target Hamiltonian $H_t$ satisfy Eq. (\ref{eq:QSA strength time relation}) which is $ t g= t^{\prime} g^{\prime}$. We also mentioned that QSA will not produce any error by itself since they are analytically exact, where errors can be from experimental imperfections. In this section, we will further discuss the strength and errors of QSA itself as well as QSA applications on the toric code and its modifications. In particular, we focus on the strength and errors of QSA for four-body interactions and general QSA, as well as the strength and errors for simulating the toric code Hamiltonian, since they cover most of the contents in this paper. Others are slight modifications of the cases we focused on and thus have similar strength and errors with those cases. For example, The Hamiltonian of the toric code with twists \cite{twist} is the toric code Hamiltonian with several five-body interactions simulated by general QSA, where its strength and errors can be described by strength and errors of the toric code Hamiltonian and the general QSA.

\subsection{Strength}

All discussions for the strength of QSA are under the same framework, which is Eq. (\ref{eq:QSA strength time relation}). In this section, we assume the time consumption between time evolutions of controlled and target Hamiltonians are the same so that we can focus on their strength difference. As briefly discussed in Sec. \ref{sec:Quantum Simulation by Attachment}, we expect that the strength difference is small for practical purposes. In this subsection, we will show that it is indeed true. The strength of the four-body QSA will be explicitly calculated. The strength of the general QSA and the toric code Hamiltonian are then calculated as two different modifications of the strength of the four-body QSA.

We now start with constructing an expression for the strength of the four-body QSA. As we can see from Eq. (\ref{eq:toric QSA four}) in Sec. \ref{sec:Simulating the Toric Code Hamiltonian}, the QSA for four-body interactions can be written as
\be
\begin{aligned}
&e^{-i t^{\prime} g^{\prime} X_1 Z_2 Z_3 X_4} \\
=& e^{-i \tau \Omega[\left(X_1 X_3+Z_3\right)+\left(X_2 X_4+Z_2\right)]}e^{-i t g X_2 X_3} \\
& \times e^{-i \tau^{\prime} \Omega^{\prime} [\left(X_1 X_3+Z_3\right)+\left(X_2 X_4+Z_2\right)]}
\end{aligned}
\label{eq:se s four expression}
\ee
where we assume that the same parameters are employed for both attachment Hamiltonians as $\tau_2 \Omega_2=\tau_3 \Omega_3=\tau \Omega$ and $\tau_2^{\prime} \Omega_2^{\prime}=\tau_3^{\prime} \Omega_3^{\prime}=\tau^{\prime} \Omega^{\prime}$. It is usually a valid assumption since two attachment Hamiltonians are identical except acting on different spins. Since controlled and target time evolutions have the same time consumption, we can obtain the relation $t^{\prime}=t+\tau+\tau^{\prime}$ for the four-body QSA from Eq. \eqref{eq:se s four expression}. Thus, given Eq. (\ref{eq:QSA strength time relation}), we can obtain the relation between the strength of original and target Hamiltonians as
\be
g^{\prime}=g \frac{t}{t+\tau+\tau^{\prime}}
\label{eq:se s strength relation with time}
\ee
which explicitly shows that the strength of the target Hamiltonian increases with decreasing the time consumption of attachment propagators. In Sec. \ref{sec:Simulating the Toric Code Hamiltonian}, we mentioned that $\tau \Omega=-\frac{ \pi}{2}$ and $\tau^{\prime} \Omega^{\prime}=\frac{ \pi}{2}$, which state the time consumption of attachment propagators decreases with increasing the strength of attachment Hamiltonians. We can also rewrite Eq. \eqref{eq:se s strength relation with time} as
\be
g^{\prime}=g \frac{t}{t-\frac{\pi}{2\Omega}+\frac{\pi}{2\Omega^{\prime}}}
\label{eq:se s strength relation with strength}
\ee
where $\Omega <0$ and $\Omega^{\prime} >0$ since both $\tau$ and $\tau^{\prime}$ must be non-negative. As one can read from Eq. (\ref{eq:se s strength relation with time}), the difference between the strength of controlled Hamiltonians and the target Hamiltonian is normally not more than one order of magnitude, since $\tau$ and $\tau^{\prime}$ is usually not much larger than $t$. This strength difference is sufficiently small for most practical purposes

Generalizing Eq. (\ref{eq:se s strength relation with time}) and Eq. (\ref{eq:se s strength relation with strength}) to strength equations of general QSA is straightforward. In Sec. \ref{sec:Quantum Simulation by Attachment}, we showed that all attachment Hamiltonians in each step of general QSA commute with each other, which means that those attachment Hamiltonians can be driven simultaneously like in the case of Eq. (\ref{eq:se s four expression}). The total time consumption of the target Hamiltonian in general QSA is given by $t^{\prime}=t+n(\tau+\tau^{\prime})$ where $n$ is the number of steps. Thus, Eq. (\ref{eq:se s strength relation with time}) and Eq. (\ref{eq:se s strength relation with strength}) is generalized to
\be
g^{\prime}=g \frac{t}{t+n(\tau+\tau^{\prime})}
\label{eq:se s strength relation with time n body}
\ee
and
\be
g^{\prime}=g \frac{t}{t+n(\frac{\pi}{2\Omega^{\prime}}-\frac{\pi}{2\Omega})}
\label{eq:se s strength relation with strength n body}
\ee
correspondingly. As mentioned in Sec. \ref{sec:Quantum Simulation by Attachment}, with a linearly increasing number of steps $n$, the number of bodies $N$ in the resulting Pauli string of general QSA increases exponentially in the best case as $N=2^{n+1}$ or increase linearly in the worst case as $N=n+2$. The number of steps $n$ for most practical purposes is not more than $10$, so the strength difference between controlled Hamiltonians and the target Hamiltonian in general QSA is usually not more than one order of magnitude, which is fairly small as the number of bodies $N$ can be very large like $N=2^{10+1}=2048$ for 10 steps.

We can also calculate the strength of the whole toric code lattice Hamiltonian using Eq. (\ref{eq:se s strength relation with time}) and Eq. (\ref{eq:se s strength relation with strength}). As introduced in Sec. \ref{sec:Simulating the Toric Code Hamiltonian}, the whole toric code lattice is divided into four groups $G_\alpha$ of four-body interactions where all four-body interactions in the same group commute with each other. As shown in Eq. (\ref{eq:toric QSA digital target}), the product of time consumption and strength is the same for the whole toric code Hamiltonian and individual four-body interactions. It can be written as $t_{w} g_{w}=t^{\prime} g^{\prime}$, where $t_{w} g_{w}$ is time consumption and strength of the whole toric code Hamiltonian and $t^{\prime} g^{\prime}$ is time consumption and strength of individual four-body interactions. Since all four-body interactions in the same group $G_\alpha$ can be simulated using QSA simultaneously and the whole toric code can be simulated in four steps where one group is simulated in one step, the time consumption of the whole toric code Hamiltonian and individual four-body interactions satisfies the relation $t_{w}=4t^{\prime}$, which implies $g_{w}=g^{\prime}/4$. Therefore, using Eq. (\ref{eq:se s strength relation with time}) and Eq. (\ref{eq:se s strength relation with strength}), we can obtain the strength relation between controlled Hamiltonians, which are two-body and single-body Hamiltonians, and the whole toric code Hamiltonian as
\be
g_{w}=g \frac{t}{4(t+\tau+\tau^{\prime})}
\label{eq:se s strength relation with time toric code}
\ee
and
\be
g_{w}=g \frac{t}{4(t-\frac{\pi}{2\Omega}+\frac{\pi}{2\Omega^{\prime}})} .
\label{eq:se s strength relation with strength toric code}
\ee
The strength difference between controlled Hamiltonians and the toric code Hamiltonian is also not more than one order of magnitude, which is sufficiently small for most practical purposes.

\subsection{Errors}

Everything introduced in this paper is analytically exact, including QSA itself as well as its application on the toric code and modified toric codes. It means there is no error in theory. However, we are not able to set parameters to be exactly the same as the theoretical requirements in an experiment, which will introduce some errors. Whether or not those errors will damage the results of QSA becomes a question. Now, we will explicitly calculate those errors to show that those errors are indeed ignorable. Like before, we will start with errors of four-body QSA, and then generalize to the cases of general QSA and QSA for simulating the toric code Hamiltonian.

For all three cases, parameter errors can be on both time $\delta t$ and strength $\delta g$, but those two kinds of errors can be combined to be a single parameter error $\delta (t g)$ since in QSA the time and strength always appear as a pair as
\be
\begin{aligned}
&(t+\delta t)(g+\delta g)\\
=&tg+\delta (tg)+\mathcal{O}(\delta^2).
\end{aligned}
\label{eq:se e parameter error combine}
\ee
We assume this kind of parameter error for all controlled Hamiltonians has a similar small size, and we label them all with a single $\delta$ symbol instead of $\delta (t g)$. The resulting errors are considered to be small enough if their size is $\mathcal{O}(\delta)$ or smaller.

In order to calculate errors of four-body QSA, we can insert those $\delta$ errors into Eq. (\ref{eq:se s four expression}) to check if the resulting error is small enough as
\be
\begin{aligned}
& e^{-i (\tau \Omega+\delta)H_A}e^{-i (t g+\delta) X_2 X_3} e^{-i (\tau^{\prime} \Omega^{\prime}+\delta) H_A}\\
=& e^{-i \delta H_A}e^{-i \tau \Omega H_A}e^{-i (t g+\delta) X_2 X_3} e^{-i \tau^{\prime} \Omega^{\prime} H_A}e^{-i \delta H_A}\\
=& e^{-i \delta H_A}e^{-i (t g+\delta) X_1 Z_2 Z_3 X_4} e^{-i \delta H_A}\\
\end{aligned}
\label{eq:se e four first step}
\ee
where $(X_1 X_3+Z_3)+( X_2 X_4+Z_2)$ in the original Eq. (\ref{eq:se s four expression}) is rewritten as $H_A$ to simplify the expression. Furthermore, since $\exp(-i\delta H)=1-i\delta H+\mathcal{O}(\delta^2)$, we can simply Eq. (\ref{eq:se e four first step}) as
\be
\begin{aligned}
& e^{-i \delta H_A}e^{-i (t g+\delta) X_1 Z_2 Z_3 X_4} e^{-i \delta H_A}\\
=&(1-i\delta H_A)e^{-i t g X_1 Z_2 Z_3 X_4}(1-i\delta X_1 Z_2 Z_3 X_4)\\
&\times (1-i\delta H_A)+\mathcal{O}(\delta^2) \\
=&e^{-i t g X_1 Z_2 Z_3 X_4}-i\delta (H_A e^{-i t g X_1 Z_2 Z_3 X_4}\\
&+e^{-i t g X_1 Z_2 Z_3 X_4}X_1 Z_2 Z_3 X_4\\
&\quad +e^{-i t g X_1 Z_2 Z_3 X_4}H_A)+\mathcal{O}(\delta^2) .
\end{aligned}
\label{eq:se e four last step}
\ee
The difference between Eq. (\ref{eq:se e four last step}) and Eq. (\ref{eq:se s four expression}) is the error of four-body QSA, which is at order $\delta$ and thus is small enough to be ignored.

Generalizing Eq. (\ref{eq:se e four first step}) and Eq. (\ref{eq:se e four last step}) to the case of general QSA is straightforward. We will analyze the error of general QSA using mathematical induction since general QSA is done by iterating QSA as introduced in Sec. \ref{sec:Quantum Simulation by Attachment}. For each iterating step of general QSA, we assume that there is an error of order $\delta$ on the original Hamiltonian propagator having the form as $\exp(-I t g H_O)+\mathcal{O}(\delta)$. Another way to state this assumption is that the previous step of any iterating step produces an error of order $\delta$. One can show that under this assumption the order of the resulting error in each step is not larger than $\delta$ as
\be
\begin{aligned}
&e^{-i \tau \Omega H_A}(e^{-i t g H_O}+\mathcal{O}(\delta)) e^{-i \tau^{\prime} \Omega^{\prime}  H_A}\\
=&e^{-i \tau \Omega H_A}e^{-i t g H_O} e^{-i \tau^{\prime} \Omega^{\prime}  H_A}+\mathcal{O}(\delta).
\end{aligned}
\label{eq:se e general error separation}
\ee
At the same time, one may notice that Eq. (\ref{eq:se e four first step}) and Eq. (\ref{eq:se e four last step}) do not rely on the form of the original Hamiltonian since their derivation process is also valid if we use an N-body interaction instead of $ X_2 X_3$ as the original Hamiltonian, so they are valid for any step of the general QSA. The first term in the last line of Eq. (\ref{eq:se e general error separation}) is a regular step of the general QSA, so Eq. (\ref{eq:se e four first step}) and Eq. (\ref{eq:se e four last step}) states that it will produce an error of order $\delta$ due to parameter imperfection. This error from parameter imperfection can be combined with the resulting error in Eq. \eqref{eq:se e general error separation} to form an overall error, where the order of overall error is also $\delta$. It means that for any iterating step, if we assume there is an error of order $\delta$ produced by the previous step, the current step will produce an error of order $\delta$. The initial step of general QSA is actually the four-body interaction QSA, which produces an error of order $\delta$. Therefore, by mathematical induction, the error of general QSA is at order $\delta$, which is sufficiently small to be ignored.

As shown in Eq. (\ref{eq:toric QSA digital target}), the whole toric code Hamiltonian is simulated by a product of many four-body interactions $P_{i, j}$. Each four-body interaction is simulated by four-body QSA which brings an error with order $\delta$, where the error of four-body QSA in Eq. \eqref{eq:se e four last step} can be relabeled as $\delta U_{i, j}$. Therefore, the error of the whole toric code Hamiltonian is
\be
\prod_{i,j} (e^{i J \tau P_{i, j}}+\delta U_{i, j})=\prod_{i,j}e^{i J \tau P_{i, j}}+\mathcal{O}(\delta)
\label{eq:se e toric error result}
\ee
which is derived by directly expanding the left-hand side of the equation and combining all terms with order $\delta$ or higher. Therefore, the error of simulating the toric code Hamiltonian using QSA is at order $\delta$, which is small enough to be ignored.

\section{Outlook}
\label{sec:Outlook}

In this paper, we introduced an analytically exact method, named quantum simulation by attachment, for simulating N-body interactions using untunable decentralized controlled Hamiltonians. QSA can be employed to simulate the toric code and its modifications. In particular, QSA can generate a superposition of altered and unaltered states of anyons, which enable the toric code quantum memory \cite{Pbook} and the toric code with holes \cite{hole} to be simulated. Since the QSA is non-perturbative and fast, the strength of simulation results is high. QSA is error-free in theory as it is analytically exact, whereas errors from experimental parameter imperfection are also shown to be ignorable.

We use QSA to simulate Pauli strings in this paper, but in fact, QSA can be further generalized to simulate Hamiltonians other than Pauli strings. As the name implies, quantum simulation by attachment is actually a method of attaching a Hamiltonian to another Hamiltonian, where those two Hamiltonians are not necessarily Pauli strings. Instead of Eq. (\ref {eq:QSA three control}), we can redefine the attachment Hamiltonian $H_{A}$ and the original Hamiltonian $H_{O}$ as
\be
\begin{aligned}
& H_{A}=\frac{1}{\sqrt{2}}\left(H_{\alpha} \otimes \mathbb{I}+H_{\beta} \otimes H_3\right) \, , \\
& H_{O}=H_1 \otimes H_{\alpha \, \text {or} \, \beta}.
\end{aligned}
\label{eq:outlook definition Hamiltonians}
\ee
General connectors $H_\alpha$ and $H_\beta$ are Hamiltonians that satisfy the Hermitian relation $H^{\dagger}_i=H_i$ and the anticommutation relation $\left\{H_i, H_j\right\}=2 \delta_{i j} \mathbb{I}$, where $i$ and $j$ representing either $\alpha$ or $\beta $. Hamiltonians $H_1$ and $H_3$ need to satisfy conditions $H^{\dagger}_3=H_3$, $H_3H_3=\mathbb{I}$, and commutation relation $\left[H_1, H_3\right]=0$. If all constraints given above are satisfied, all results of the three-body QSA in Sec. \ref{sec:Quantum Simulation by Attachment} are also valid for redefined $H_{A}$ and $H_{O}$ as
\be
e^{-i \tau \lambda H_A}=\left\{\begin{array}{l}
U_{A}=\frac{i}{\sqrt{2}}\left(H_{\alpha} \otimes \mathbb{I}+H_{\beta} \otimes H_3\right),\\ 
\quad \text {if }  \lambda \tau=\frac{3 \pi}{2}+2 \pi m, \, m \in \mathbb{Z} \\
U^{\dagger}_{A}=\frac{-i}{\sqrt{2}}\left(H_{\alpha} \otimes \mathbb{I}+H_{\beta} \otimes H_3\right),\\ 
\quad \text {if } \lambda \tau=\frac{\pi}{2}+2 \pi m^{\prime},\, m^{\prime} \in \mathbb{Z}
\end{array}\right. 
\label{eq:outlook definition Hamiltonians propagator}
\ee
and
\be
U_{A}H_{O}U^{\dagger}_{A}=H_1 \otimes H_{k+1} \otimes H_3 
\label{eq:outlook definition Hamiltonians QSA}
\ee
where $k+1$ also represents another connector like in Sec. \ref{sec:Quantum Simulation by Attachment}.

General QSA in Sec. \ref{sec:Quantum Simulation by Attachment} can also be done using those redefined $H_{A}$ and $H_{O}$. If there is always at least one connector $H_\alpha$ or $H_\beta$ in the original Hamiltonian of each iterating step in general QSA, the general QSA with redefined $H_{A}$ and $H_{O}$ is valid since each iterating step can be implemented by the modified three-body QSA introduced above. We can prove the existence of at least one connector in the original Hamiltonian of each iterating step by mathematical induction. In each step, if we assume that there is at least one connector in the original Hamiltonian, the target Hamiltonian of this step will contain at least one connector because $H_{A}$ is attached to $H_{O}$ by swapping connectors where the swapped connector itself is also a valid connector. The target Hamiltonian of this step is the original Hamiltonian of the next step, so there is at least one connector in the original Hamiltonian of the next step if there is at least one connector in the original Hamiltonian of the current step. We can always artificially construct the initial step such that there is at least one connector in the original Hamiltonian of the initial step. Therefore, by mathematical induction, there is at least one connector in the original Hamiltonian of each iterating step. It implies that the general QSA with redefined $H_{A}$ and $H_{O}$ is valid. Those modified three-body QSA and general QSA enable us to simulate a much more general target Hamiltonian as well as to use a broader range of controlled Hamiltonians.

Besides those modified QSA, the regular QSA also has many further applications. Firstly, the fact that one is able to simulate N-body, instead of only four-body, Pauli string using QSA enables more models other than the toric code to be simulated by QSA. Many models of artificial quantum systems, like Kitaev’s honeycomb lattice model \cite{KitaevHoneycomb, Pbook} used in topological quantum computation, are based on Pauli strings. If a model is based on Pauli strings, there is a chance that QSA can be employed to simulate this model. Secondly, as mentioned in Sec. \ref{sec:Ground State and Anyons of the Toric Code}, if anyons are braided by Pauli string, one can use QSA to generate a superposition of braided and unbraided states. In some topological quantum computation models, this ability enables us to perform more kinds of operations since some superposition of states is not obtainable without QSA. Besides that, some operations can be performed faster, since a wider range of resulting states can be obtained in each step which could decrease the number of steps required to reach a particular state. We have demonstrated the power of this ability in the toric code quantum memory \cite{Pbook} and the toric code with holes \cite{hole}. This ability is not normally assumed in a topological quantum computation model as it may impose some intrinsic errors from experimental parameter imperfection, as discussed in Sec. \ref{sec:Strength and Errors}. However, this ability is still valuable since errors are usually small. Further research can be done on how to employ this ability on more topological quantum computation models to enable those models to perform a broader range of computations or perform computations faster.

Besides those direct applications of QSA, other further research can also be done on this topic. For example, one may try to extend the range of target Hamiltonians obtainable by QSA. The regular QSA itself can only simulate a single Pauli string, whereas a summation of Pauli strings must be simulated by other approaches. If one can find a method to simulate a summation of Pauli strings from individual Pauli strings, one can significantly extend the ability of QSA such that one is able to or close to simulate a general Hamiltonian from two-body Pauli strings.

\bibliography{bibfile.bib}

\begin{thebibliography}{14}%
\makeatletter
\providecommand \@ifxundefined [1]{%
 \@ifx{#1\undefined}
}%
\providecommand \@ifnum [1]{%
 \ifnum #1\expandafter \@firstoftwo
 \else \expandafter \@secondoftwo
 \fi
}%
\providecommand \@ifx [1]{%
 \ifx #1\expandafter \@firstoftwo
 \else \expandafter \@secondoftwo
 \fi
}%
\providecommand \natexlab [1]{#1}%
\providecommand \enquote  [1]{``#1''}%
\providecommand \bibnamefont  [1]{#1}%
\providecommand \bibfnamefont [1]{#1}%
\providecommand \citenamefont [1]{#1}%
\providecommand \href@noop [0]{\@secondoftwo}%
\providecommand \href [0]{\begingroup \@sanitize@url \@href}%
\providecommand \@href[1]{\@@startlink{#1}\@@href}%
\providecommand \@@href[1]{\endgroup#1\@@endlink}%
\providecommand \@sanitize@url [0]{\catcode `\\12\catcode `\$12\catcode `\&12\catcode `\#12\catcode `\^12\catcode `\_12\catcode `\%12\relax}%
\providecommand \@@startlink[1]{}%
\providecommand \@@endlink[0]{}%
\providecommand \url  [0]{\begingroup\@sanitize@url \@url }%
\providecommand \@url [1]{\endgroup\@href {#1}{\urlprefix }}%
\providecommand \urlprefix  [0]{URL }%
\providecommand \Eprint [0]{\href }%
\providecommand \doibase [0]{https://doi.org/}%
\providecommand \selectlanguage [0]{\@gobble}%
\providecommand \bibinfo  [0]{\@secondoftwo}%
\providecommand \bibfield  [0]{\@secondoftwo}%
\providecommand \translation [1]{[#1]}%
\providecommand \BibitemOpen [0]{}%
\providecommand \bibitemStop [0]{}%
\providecommand \bibitemNoStop [0]{.\EOS\space}%
\providecommand \EOS [0]{\spacefactor3000\relax}%
\providecommand \BibitemShut  [1]{\csname bibitem#1\endcsname}%
\let\auto@bib@innerbib\@empty
\bibitem [{\citenamefont {Pachos}(2012)}]{Pbook}%
  \BibitemOpen
  \bibfield  {author} {\bibinfo {author} {\bibfnamefont {J.~K.}\ \bibnamefont {Pachos}},\ }\href {https://doi.org/DOI: 10.1017/CBO9780511792908} {\emph {\bibinfo {title} {Introduction to Topological Quantum Computation}}}\ (\bibinfo  {publisher} {Cambridge University Press},\ \bibinfo {address} {Cambridge},\ \bibinfo {year} {2012})\BibitemShut {NoStop}%
\bibitem [{\citenamefont {Kitaev}(2003)}]{Ktoric}%
  \BibitemOpen
  \bibfield  {author} {\bibinfo {author} {\bibfnamefont {A.~Y.}\ \bibnamefont {Kitaev}},\ }\bibfield  {title} {\bibinfo {title} {Fault-tolerant quantum computation by anyons},\ }\href {https://doi.org/https://doi.org/10.1016/S0003-4916(02)00018-0} {\bibfield  {journal} {\bibinfo  {journal} {Annals of Physics}\ }\textbf {\bibinfo {volume} {303}},\ \bibinfo {pages} {2} (\bibinfo {year} {2003})}\BibitemShut {NoStop}%
\bibitem [{\citenamefont {Bombin}(2010)}]{twist}%
  \BibitemOpen
  \bibfield  {author} {\bibinfo {author} {\bibfnamefont {H.}~\bibnamefont {Bombin}},\ }\bibfield  {title} {\bibinfo {title} {Topological order with a twist: Ising anyons from an abelian model},\ }\href {https://doi.org/10.1103/PhysRevLett.105.030403} {\bibfield  {journal} {\bibinfo  {journal} {Physical Review Letters}\ }\textbf {\bibinfo {volume} {105}},\ \bibinfo {pages} {030403} (\bibinfo {year} {2010})}\BibitemShut {NoStop}%
\bibitem [{\citenamefont {Wootton}(2012)}]{hole}%
  \BibitemOpen
  \bibfield  {author} {\bibinfo {author} {\bibfnamefont {J.~R.}\ \bibnamefont {Wootton}},\ }\bibfield  {title} {\bibinfo {title} {Quantum memories and error correction},\ }\href {https://doi.org/10.1080/09500340.2012.737937} {\bibfield  {journal} {\bibinfo  {journal} {Journal of Modern Optics}\ }\textbf {\bibinfo {volume} {59}},\ \bibinfo {pages} {1717} (\bibinfo {year} {2012})}\BibitemShut {NoStop}%
\bibitem [{\citenamefont {Nayak}\ \emph {et~al.}(2008)\citenamefont {Nayak}, \citenamefont {Simon}, \citenamefont {Stern}, \citenamefont {Freedman},\ and\ \citenamefont {Das~Sarma}}]{RMPreviewTQC}%
  \BibitemOpen
  \bibfield  {author} {\bibinfo {author} {\bibfnamefont {C.}~\bibnamefont {Nayak}}, \bibinfo {author} {\bibfnamefont {S.~H.}\ \bibnamefont {Simon}}, \bibinfo {author} {\bibfnamefont {A.}~\bibnamefont {Stern}}, \bibinfo {author} {\bibfnamefont {M.}~\bibnamefont {Freedman}},\ and\ \bibinfo {author} {\bibfnamefont {S.}~\bibnamefont {Das~Sarma}},\ }\bibfield  {title} {\bibinfo {title} {Non-abelian anyons and topological quantum computation},\ }\href {https://doi.org/10.1103/RevModPhys.80.1083} {\bibfield  {journal} {\bibinfo  {journal} {Reviews of Modern Physics}\ }\textbf {\bibinfo {volume} {80}},\ \bibinfo {pages} {1083} (\bibinfo {year} {2008})}\BibitemShut {NoStop}%
\bibitem [{\citenamefont {Petiziol}\ \emph {et~al.}(2024)\citenamefont {Petiziol}, \citenamefont {Wimberger}, \citenamefont {Eckardt},\ and\ \citenamefont {Mintert}}]{hybrid4}%
  \BibitemOpen
  \bibfield  {author} {\bibinfo {author} {\bibfnamefont {F.}~\bibnamefont {Petiziol}}, \bibinfo {author} {\bibfnamefont {S.}~\bibnamefont {Wimberger}}, \bibinfo {author} {\bibfnamefont {A.}~\bibnamefont {Eckardt}},\ and\ \bibinfo {author} {\bibfnamefont {F.}~\bibnamefont {Mintert}},\ }\bibfield  {title} {\bibinfo {title} {Nonperturbative floquet engineering of the toric-code hamiltonian and its ground state},\ }\href {https://doi.org/10.1103/PhysRevB.109.075126} {\bibfield  {journal} {\bibinfo  {journal} {Physical Review B}\ }\textbf {\bibinfo {volume} {109}},\ \bibinfo {pages} {075126} (\bibinfo {year} {2024})}\BibitemShut {NoStop}%
\bibitem [{\citenamefont {Petiziol}\ \emph {et~al.}(2021)\citenamefont {Petiziol}, \citenamefont {Sameti}, \citenamefont {Carretta}, \citenamefont {Wimberger},\ and\ \citenamefont {Mintert}}]{hybrid3}%
  \BibitemOpen
  \bibfield  {author} {\bibinfo {author} {\bibfnamefont {F.}~\bibnamefont {Petiziol}}, \bibinfo {author} {\bibfnamefont {M.}~\bibnamefont {Sameti}}, \bibinfo {author} {\bibfnamefont {S.}~\bibnamefont {Carretta}}, \bibinfo {author} {\bibfnamefont {S.}~\bibnamefont {Wimberger}},\ and\ \bibinfo {author} {\bibfnamefont {F.}~\bibnamefont {Mintert}},\ }\bibfield  {title} {\bibinfo {title} {Quantum simulation of three-body interactions in weakly driven quantum systems},\ }\href {https://doi.org/10.1103/PhysRevLett.126.250504} {\bibfield  {journal} {\bibinfo  {journal} {Physical Review Letters}\ }\textbf {\bibinfo {volume} {126}},\ \bibinfo {pages} {250504} (\bibinfo {year} {2021})}\BibitemShut {NoStop}%
\bibitem [{\citenamefont {Müller}\ \emph {et~al.}(2011)\citenamefont {Müller}, \citenamefont {Hammerer}, \citenamefont {Zhou}, \citenamefont {Roos},\ and\ \citenamefont {Zoller}}]{nbodyS}%
  \BibitemOpen
  \bibfield  {author} {\bibinfo {author} {\bibfnamefont {M.}~\bibnamefont {Müller}}, \bibinfo {author} {\bibfnamefont {K.}~\bibnamefont {Hammerer}}, \bibinfo {author} {\bibfnamefont {Y.~L.}\ \bibnamefont {Zhou}}, \bibinfo {author} {\bibfnamefont {C.~F.}\ \bibnamefont {Roos}},\ and\ \bibinfo {author} {\bibfnamefont {P.}~\bibnamefont {Zoller}},\ }\bibfield  {title} {\bibinfo {title} {Simulating open quantum systems: from many-body interactions to stabilizer pumping},\ }\href {https://doi.org/10.1088/1367-2630/13/8/085007} {\bibfield  {journal} {\bibinfo  {journal} {New Journal of Physics}\ }\textbf {\bibinfo {volume} {13}},\ \bibinfo {pages} {085007} (\bibinfo {year} {2011})}\BibitemShut {NoStop}%
\bibitem [{\citenamefont {Wen}(2003)}]{Wtoric}%
  \BibitemOpen
  \bibfield  {author} {\bibinfo {author} {\bibfnamefont {X.-G.}\ \bibnamefont {Wen}},\ }\bibfield  {title} {\bibinfo {title} {Quantum orders in an exact soluble model},\ }\href {https://doi.org/10.1103/PhysRevLett.90.016803} {\bibfield  {journal} {\bibinfo  {journal} {Physical Review Letters}\ }\textbf {\bibinfo {volume} {90}},\ \bibinfo {pages} {016803} (\bibinfo {year} {2003})}\BibitemShut {NoStop}%
\bibitem [{\citenamefont {Dennis}\ \emph {et~al.}(2002)\citenamefont {Dennis}, \citenamefont {Kitaev}, \citenamefont {Landahl},\ and\ \citenamefont {Preskill}}]{DennisQuantumMemory}%
  \BibitemOpen
  \bibfield  {author} {\bibinfo {author} {\bibfnamefont {E.}~\bibnamefont {Dennis}}, \bibinfo {author} {\bibfnamefont {A.}~\bibnamefont {Kitaev}}, \bibinfo {author} {\bibfnamefont {A.}~\bibnamefont {Landahl}},\ and\ \bibinfo {author} {\bibfnamefont {J.}~\bibnamefont {Preskill}},\ }\bibfield  {title} {\bibinfo {title} {Topological quantum memory},\ }\href@noop {} {\bibfield  {journal} {\bibinfo  {journal} {Journal of Mathematical Physics}\ }\textbf {\bibinfo {volume} {43}},\ \bibinfo {pages} {4452} (\bibinfo {year} {2002})}\BibitemShut {NoStop}%
\bibitem [{\citenamefont {Bravyi}(2006)}]{isinguniversal}%
  \BibitemOpen
  \bibfield  {author} {\bibinfo {author} {\bibfnamefont {S.}~\bibnamefont {Bravyi}},\ }\bibfield  {title} {\bibinfo {title} {Universal quantum computation with the $\ensuremath{\nu}=5/2$ fractional quantum hall state},\ }\href {https://doi.org/10.1103/PhysRevA.73.042313} {\bibfield  {journal} {\bibinfo  {journal} {Physical Review A}\ }\textbf {\bibinfo {volume} {73}},\ \bibinfo {pages} {042313} (\bibinfo {year} {2006})}\BibitemShut {NoStop}%
\bibitem [{\citenamefont {Bravyi}\ and\ \citenamefont {Kitaev}(2005)}]{magic1}%
  \BibitemOpen
  \bibfield  {author} {\bibinfo {author} {\bibfnamefont {S.}~\bibnamefont {Bravyi}}\ and\ \bibinfo {author} {\bibfnamefont {A.}~\bibnamefont {Kitaev}},\ }\bibfield  {title} {\bibinfo {title} {Universal quantum computation with ideal clifford gates and noisy ancillas},\ }\href {https://doi.org/10.1103/PhysRevA.71.022316} {\bibfield  {journal} {\bibinfo  {journal} {Physical Review A}\ }\textbf {\bibinfo {volume} {71}},\ \bibinfo {pages} {022316} (\bibinfo {year} {2005})}\BibitemShut {NoStop}%
\bibitem [{\citenamefont {Knill}(2004)}]{magic2}%
  \BibitemOpen
  \bibfield  {author} {\bibinfo {author} {\bibfnamefont {E.}~\bibnamefont {Knill}},\ }\bibfield  {title} {\bibinfo {title} {Fault-tolerant postselected quantum computation: Threshold analysis},\ }\href@noop {} {\bibfield  {journal} {\bibinfo  {journal} {arXiv preprint quant-ph/0404104}\ } (\bibinfo {year} {2004})}\BibitemShut {NoStop}%
\bibitem [{\citenamefont {Kitaev}(2006)}]{KitaevHoneycomb}%
  \BibitemOpen
  \bibfield  {author} {\bibinfo {author} {\bibfnamefont {A.}~\bibnamefont {Kitaev}},\ }\bibfield  {title} {\bibinfo {title} {Anyons in an exactly solved model and beyond},\ }\href@noop {} {\bibfield  {journal} {\bibinfo  {journal} {Annals of Physics}\ }\textbf {\bibinfo {volume} {321}},\ \bibinfo {pages} {2} (\bibinfo {year} {2006})}\BibitemShut {NoStop}%
\end{thebibliography}%

\end{document}